\documentstyle[multicol,aps,pre,twocolumn,epsfig]{revtex}

\begin{document}
\twocolumn[\hsize\textwidth\columnwidth\hsize\csname@twocolumnfalse\endcsname
\title{Fronts in passive scalar turbulence}
\author{A.~Celani$^{1,2}$, A.~Lanotte$^{2}$, A.~Mazzino$^{3}$ 
and M.~Vergassola$^{2}$ \\
\small{$^{1}$ CNRS, INLN, 1361 Route des Lucioles,
06560 Valbonne, France.}\\
\small{$^{2}$ CNRS, Observatoire de la C\^ote d'Azur, B.P. 4229,
06304 Nice Cedex 4, France.}\\
\small{$^{3}$ INFM--Dipartimento di Fisica, Universit\`a di Genova,
Via Dodecaneso, 33, I-16142 Genova, Italy.}}
\date{\today}
\maketitle
\begin{abstract}
The evolution of scalar fields transported by turbulent flow is
characterized by the presence of fronts, which rule the small-scale
statistics of scalar fluctuations. With the aid of numerical
simulations, it is shown that\,: isotropy is not recovered, in the
classical sense, at small scales; scaling exponents are universal with
respect to the scalar injection mechanisms\,; high-order exponents
saturate to a constant value\,; non-mature fronts dominate the
statistics of intense fluctuations. Results on the statistics inside
the ``plateaux'', where fluctuations are weak, are also presented.
Finally, we analyze the statistics of scalar dissipation and scalar
fluxes.
\end{abstract}
]
\section{Introduction}
The understanding of small-scale fluctuations in scalar fields, such
as temperature, pollutant density, chemical or biological species
concentration, advected by turbulent flow is of great interest in both
theoretical and practical domains \cite{SS99}. Propagation phenomena
of light beams and radio waves in the atmosphere are for example
strongly influenced both by the magnitude and the spatial distribution
of small-scale temperature gradients \cite{DSCV94}. Their dynamical
properties have been a subject of very accurate experimental
investigations carried out in the last few years both in the
atmosphere \cite{DSCV94,GGZN85}, in the ocean \cite{G80,L88} and for
laboratory turbulent flow \cite{CG77,PGM82,KRS91,MW98}. A striking
feature of all these situations is the presence of fronts (also called
sheets or cliffs). The important point is that the scalar field has
very strong variations across the fronts, separated by large regions
(``ramps'' or ``plateaux'') where scalar fluctuations are weak (see
for example Fig.~\ref{r-and-c}).

In the atmosphere, ``ramp-and-cliff'' structures in the temperature
field extend from the ground to the middle stratosphere
\cite{DSCV94}. They are of major importance for the ubiquitous
stratospheric radar echoes \cite{SS92} and the problem of their
formation is known as frontogenesis \cite{P87}.  For the temperature
field it is crucial to take into account its active nature, i.e. the
fact that it influences the advecting velocity field.  Simplified
quasi and/or semi-geostrophic models have been proposed and
investigated for the frontogenesis and we refer the interested reader
to Refs.~\cite{P87,CMT94} and references therein.

Cliffs such as those observed in the active case have in fact been
observed also for passive situations, i.e. when the scalar field does
not affect the advecting velocity field.  Remarkably, cliffs are not
mere footprints of velocity structures and they arise both for
velocity fields that are solutions of the Navier--Stokes (NS)
equations \cite{P94,CLMV99} and for synthetic velocity fields
\cite{HS94,VM97,FGLP97,CK98}, including when the velocity is rapidly
varying in time (the so-called Kraichnan case \cite{K94}).  As
emphasized in Refs.~\cite{P94,HS94,W91}, the velocity gradient
stretching is weak in the elliptic regions of the velocity field,
where their `rotational' character inhibits the formation of strong
scalar gradients. Conversely, compressions along specific directions
take place in the hyperbolic regions of the incompressible flow,
scalar particles coming from distant regions of space approach each
other and strong scalar gradients are thus developed.

The observation of ramp-and-cliff structures naturally raises the
issue of their consequences for the scalar statistical properties and
their possible implications for models of scalar transport.  Two
points will be specifically addressed here for the passive scalar
case. Part of the results were previously summarized in
Ref.~\cite{CLMV99}.

The first concerns the role of ramps and cliffs for the scalar
anisotropy.  A central postulate of the classical
Kolmogorov--Obukhov--Corrsin theory (see, e.g., Ref.~\cite{T72}) for
passive scalar turbulence is that the anisotropy degree should decay
as smaller and smaller scales are considered. When the system is
driven by an anisotropic forcing, such as a large-scale scalar
gradient, the most common experimental situation, isotropy is usually
supposed to be fully recovered at small scales. However, the evidence
stemming from the experiments is that this is not the case (at least
not in the original sense, see the following). Ramps and cliffs are
preferentially aligned with the direction of the large-scale scalar
gradient and this permits the anisotropies to find their way down to
the smallest excited scales.  In particular, the probability density
function (pdf) for the scalar derivatives along the direction of the
large-scale scalar gradient are strongly skewed. Furthermore, the
scalar gradient skewness (that should vanish for an isotropic field)
is $O(1)$, independently of the P\'eclet number
\cite{CG77,PGM82,KRS91,MW98,P94,HS94}. This leads us to investigate
scalar turbulence universality, i.e. its degree of independence of
injection mechanisms.

Recent numerical and experimental works exploiting symmetries under
rotations point to the fact that anisotropic turbulent fluctuations
are subdominant with respect to isotropic ones. We shall discuss
the relation between these results and the
aforementioned experimental observations about the persistence of
large-scale anisotropies.

The second issue associated to the presence of cliffs concerns their
role for the statistics of scalar difference strong fluctuations,
i.e. to the behavior of high-order structure functions. In particular,
scalar variations across the cliffs are comparable to the
root-mean-square (rms) value of the scalar. This suggests that scalar
structure function scaling exponents might saturate, that is tend to a
constant for large enough orders. The question of saturation was first
raised in Ref.~\cite{K94}. Here, we shall provide evidence for
saturation in two different situations\,: scalar advection by a
two-dimensional Navier--Stokes velocity field in the inverse cascade
regime and in the three-dimensional Kraichnan model \cite{K94}. As
discussed in the sequel, the latter model represents the least
favorable case to observe saturation, because of the velocity short
correlation time. The fact that saturation is still observed points
to the genericity of the phenomenon for scalar turbulence.

The paper is organized as follows. In Sections ~\ref{basic} and
~\ref{models}, we recall some well-known results about the expected
classical laws for the decay of anisotropies in scalar turbulence and
the properties of the Navier-Stokes two-dimensional inverse energy
cascade, respectively. Anisotropies and scalar turbulence universality
are discussed in Section~\ref{aniso}, while the results about
saturation are presented in Section~\ref{int-satu}. The successive
Section investigates ramp-and-cliff structure dynamical processes.
The aim is to clarify whether the observed saturation arises from
`mature' cliffs, having thicknesses comparable to the dissipative
scale, or from `non-mature' ones, still in the process of
steepening. Section~\ref{plat} reports results on the scalar
statistics inside the ``plateaux''.  Scalar dissipation is the subject
of Section~\ref{dissip}, and the results on scalar fluxes, involving
joint velocity-scalar correlations, are discussed in
Section~\ref{fluxes}.  The last Section is devoted to conclusions.

\section{The classical theory of scalar turbulence}
\label{basic}
The advection-diffusion partial differential equation governing the
evolution of a passive scalar field $T(\bbox{r},t)$ is\,:
\begin{equation}
\label{fp}
\partial_t T(\bbox{r},t)+\bbox{v}(\bbox{r},t)\cdot\nabla\,
T(\bbox{r},t)=\kappa\Delta T(\bbox{r},t).
\end{equation}
Here, $\bbox{v}(\bbox{r},t)$ is the incompressible advecting velocity
field and $\kappa$ is the molecular diffusivity. The ``energy'' $T^2$
is statistically conserved by the advection term in (\ref{fp}) and
dissipated by the viscous term.  In order to attain a stationary state
it is thus necessary to inject scalar fluctuations. The simplest way
is to add a forcing term $f(\bbox{r},t)$ to the right-hand side (rhs)
of (\ref{fp}).  A convenient choice that we shall use in this work is
to take the forcing $f$ random, Gaussian, statistically homogeneous,
isotropic, white in time, of zero mean and correlation function
\begin{equation}
\langle f(\bbox{r},t)\,f(\bbox{0},0)\rangle =
\chi\left(r/L_f\right)\,\delta(t).
\label{fcorr}
\end{equation} 
The correlation $\chi\left(r/L_f\right)$ is concentrated at the
forcing integral scale $L_f$ and rapidly decreases for $r\gg L_f$.\\
An alternative injection mechanism, closer to typical experimental
situations, is to maintain a mean scalar gradient $\bbox{g}$. Scalar
fluctuations $\theta$ superimposed on it,
$T(\bbox{r},t)\equiv\theta(\bbox{r},t) + \bbox{g}\cdot \bbox{r}$, obey
the equation of motion\,:
\begin{equation}
\label{fp1}
\partial_t \theta(\bbox{r},t)+\bbox{v}(\bbox{r},t)\cdot\nabla\,
\theta(\bbox{r},t)=\kappa\Delta \theta(\bbox{r},t)-\bbox{g}\cdot 
\bbox{v} .
\end{equation}
Note that the mean gradient injection mechanism selects by its very
definition a preferential direction in the system and isotropy is thus
broken.

We shall be interested in the passive scalar structure
functions
\begin{equation}
\label{structuref}
S_n(\bbox{r})\equiv\langle[T(\bbox{x}+\bbox{r},t)-T(\bbox{x},t)]^n\rangle
\equiv \langle \left(\delta_r T\right)^n\rangle,
\end{equation}
where $\langle\,\rangle$ denotes the average over the velocity, and
possibly the forcing, ensemble. 
The reason to focus on correlations of the field $T$ rather than those of
the field $\theta$ is explained in Appendix~\ref{so2}.
Throughout this work we assume the
velocity to be scale-invariant, at least in the scalar inertial range
of scales defined by $\eta\ll |\bbox{r}| \ll L$.  Here, $\eta$ is the
scalar dissipation scale and $L$ is the scalar integral scale.  For
the mean-gradient injection, $L$ is comparable to the velocity
integral scale. For the randomly forced case, we shall consider
situations where the velocity integral scale is comparable (or even
larger) than the forcing one and therefore $L\simeq L_f$.  On account
of these assumptions, the scalar structure functions are expected to
have a scaling behavior in the inertial range of scales. For the
isotropic randomly forced case, odd orders vanish by symmetry. For the
injection by a mean gradient, both even and odd-order structure
functions are non-zero.

For even-order structure functions, the classical prediction of the
Kolmogorov--Obukhov--Corrsin (KOC) theory is\,:
\begin{equation}
\label{evenmom}
S_{2n}(r)=C_{2n}\epsilon_{\theta}^n\,\epsilon_{v}^{-n/3}
r^{2n/3} \left({L\over r}\right)^{\rho_{n}}\propto r^{\zeta_{2n}},
\end{equation}
with $\zeta_{2n}=2n/3$ and the anomaly $\rho_{2n} \equiv 2n/3-
\zeta_{2n}=0$.  In the previous formula, $\epsilon_{\theta} \equiv
\kappa \langle (\nabla\theta)^2 \rangle$ is the mean scalar energy
dissipation, $C_{2n}$ are non-dimensional constants and the velocity
is assumed to be of Kolmogorov-type, with velocity increments
$\langle(\delta_r v)^n\rangle \propto r^{n/3}$ and a finite energy
flux $\epsilon_{v}$.  The same arguments can be reformulated along the
same lines for other types of velocity fields.

Dimensional arguments are easily extended to odd-order moments in the
presence of a mean gradient $\bbox{g}$ \cite{Lu67}. The balance
between the advection term and the injection term on the rhs of
(\ref{fp1}) gives indeed
\begin{equation}
\label{oddmom}
S_{2n+1}(\bbox{r}) \sim \left(\bbox{g}\cdot\bbox{r}\right)
S_{2n}(\bbox{r}) \propto r^{\zeta_{2n+1}},
\end{equation}
with $\zeta_{2n+1}=2n/3+1$.  The classical prediction for the decay
rate of hyperskewnesses $R_n$ is then\,:
\begin{equation}
R_n\equiv \frac{S_{2n+1}}{S_{2}^{n+1/2}}\propto r^{2/3}\qquad
n=1,2,\cdots
\label{ske}
\end{equation}

Predictions for the scaling of scalar gradient moments with the
P\'eclet number are obtained by substituting $r=\eta\propto {\rm
Pe}^{-3/4}$ in the structure function expressions (\ref{evenmom}) and
(\ref{ske}). It follows in particular that, in the presence of a
large-scale gradient $\bbox{g}$, the skewness of the scalar gradient
along this direction should decrease as ${\rm Pe}^{-1/2}$.

Notice that the classical KOC scalings do not explicitly involve the
integral scale $L$. Anomalous scalings correspond to the violations of
these behaviors, i.e.  $\zeta_{n}\neq n/3$, and the appearance of the
integral scale $L$ in the structure function expressions via the
non-vanishing anomalies $\rho_n$.

\section{The advecting velocity fields}
\label{models}

We discuss now the choice of the advecting velocity field in
Eqs.~(\ref{fp}) or (\ref{fp1}). The flow that we shall consider mostly
in this work is the one arising from a two-dimensional Navier--Stokes
inverse energy cascade \cite{K67a}. Its properties are briefly
recalled in the following subsection. In order to corroborate the
results on saturation, we shall also consider in
Section~\ref{sec:sat-rhk} three-dimensional synthetic flows of the
Kraichnan type (see Refs.~\cite{K94,K68}).

\subsection{The 2D Navier-Stokes flow}
\label{sec:ns}

The two-dimensional Navier--Stokes equation for the vorticity
$\omega(\bbox{r},t)=-\Delta\psi(\bbox{r},t)$ is\,:
\begin{equation}
\label{NS2D}
\partial_t\omega+J\left(\omega,\psi\right)=\nu\Delta\omega
-\alpha\omega-\Delta {\cal F},
\end{equation}
where the velocity $\bbox{v}$ is related to the stream function $\psi$
as $\bbox{v} =\nabla^{\perp}\psi=(\nabla_y\psi,-\nabla_x\psi)$, the
Jacobian is denoted by $J$ and the forcing by ${\cal F}$. The friction
linear term $-\alpha \omega$ extracts energy from the system at scales
comparable to the friction scale $\eta_{\rm fr}\sim \epsilon_{v}^{1/2}
\alpha^{-3/2}$, assuming a Kolmogorov scaling law for the velocity. As
before, $\epsilon_{v}$ indicates the mean energy dissipation.

Let us first recall from Ref.~\cite{BCV99} some technical details on
the numerical integration of (\ref{NS2D}).  To focus on the inverse
cascade phase and avoid the Bose-Einstein condensation in the gravest
mode \cite{K67a}, we choose $\alpha$ to make $\eta_{\rm fr}$
sufficiently smaller than the box size. The other relevant length in
the problem is the small-scale forcing correlation length $l_{\cal
F}$, bounding the inertial range for the inverse cascade as $l_{\cal
F} \ll r\ll \eta_{\rm fr}$. We use a Gaussian forcing with correlation
function $\langle {\cal F}(\bbox{r},t)\,{\cal F}(\bbox{0},t')
\rangle=\delta(t-t')\,F(r/l_{\cal F})$.  The forcing correlation
decays rapidly for $r\gg l_{\cal F}$ and we choose $F(x)= F_0 l_{\cal
F}^2 \exp(-x^2/2)$, where $F_0$ is the energy injection rate. The
integration is performed by a standard $2/3$-dealiased pseudospectral
method on a doubly periodic square domain $2048\times 2048$.  The
viscous term in Eq.~(\ref{NS2D}) has the role of removing the
enstrophy at scales smaller than $l_{\cal F}$ and, as customary, it is
numerically more convenient to substitute it by a hyperviscous term
(of order eight in our simulations). Time evolution is implemented by
a standard second-order Adams-Bashforth scheme. The integration is
carried out for one hundred eddy turn-over times after both the
velocity and the scalar fields have reached the stationary state.

\begin{figure}
\epsfxsize=7.6truecm
\epsfbox{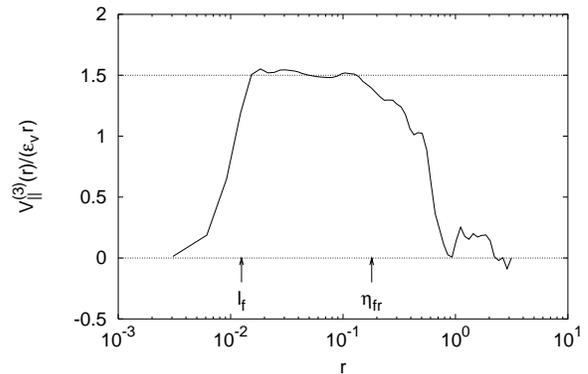}
\caption{Compensated third order longitudinal structure function 
$V^{(3)}_{\parallel}(r)/(\epsilon_{v} r)$. The dotted line is the value $3/2$. 
The vertical scale is linear.}
\label{fbis1}
\end{figure}

The inverse nature of the energy cascade clearly emerges from the
third-order longitudinal structure function
$V^{(3)}_{\parallel}(r)\equiv\langle\left[\left(\bbox{v}(\bbox{r},t)-
\bbox{v}(\bbox{0},t)\right) \cdot{\bbox{r}/r}\right]^3\rangle$.  In
analogy with the $4/5$ law valid for the three-dimensional case (see,
e.g., Ref.~\cite{UF95}), one can indeed easily derive from
(\ref{NS2D}) the relation $V^{(3)}_{\parallel}(r)=3/2\epsilon_v r$,
expressing the upward constant energy flux.  The neat plateau at the
positive value $3/2$ visible in Fig.~\ref{fbis1} is a clear
confirmation of the inverse energy cascade.

\narrowtext

\begin{figure}
\epsfxsize=7.6truecm
\epsfbox{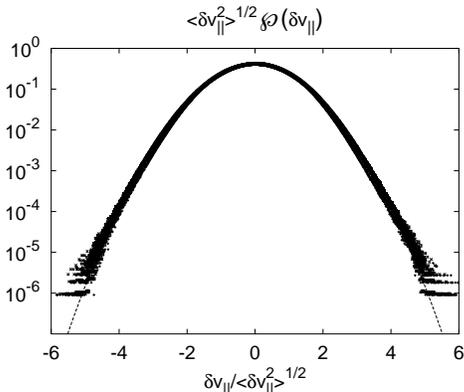}
\caption{ Pdf's of velocity fluctuations rescaled by their variance,
for five lengthscales within the inverse energy cascade range.
The dashed line is the Gaussian probability density.} 
\label{fbis2}
\end{figure}

An important property of the flow that has emerged both from numerical
simulations \cite{SY,BCV99} and experiments \cite{PT98} is the absence
of intermittency in the velocity increment statistics.  No corrections
to the Kolmogorov scaling $r^{n/3}$ for the $n$-th order velocity
structure function can be measured. In Fig.~\ref{fbis2} we show indeed
that velocity increment pdf's at different separations $r$ can all be
collapsed rescaling the increment $\delta_r v$ by their variance
$\langle (\delta_r v)^2 \rangle^{1/2} \sim r^{1/3}$.  Furthermore,
deviations from a Gaussian are small (at least for fluctuations not
very large) and this has been exploited to propose a theory for the
observed absence of intermittency \cite{VY99}.

The important point to our purposes here is that the velocity field
arising from the 2D inverse energy cascade is scale-invariant (no
intermittency) of exponent $1/3$, isotropic and, mostly important, has
realistic correlation times. Contrary to synthetic flows, like those
in \cite{HS94} or the Kraichnan ensemble used in
Section~\ref{sec:sat-rhk}, the correlation times are finite and free
of any sweeping-related pathology. The anisotropic and intermittency
properties of the scalar field that will be discussed in the sequel
are therefore intrinsic of the advection-diffusion passive scalar
dynamics and not mere footprints of the advecting velocity.

\narrowtext
\section{Anisotropy and universality}
\label{aniso}

The classical picture of the anisotropy decay presented in
Section~\ref{basic} is contradicted by the experimental
observations. Contrary to the expected ${\rm Pe}^{-1/2}$ behavior, the
skewness of the scalar derivative along the large-scale gradient
$\partial_{\|}\theta \equiv \bbox{g}\cdot \nabla \theta /|\bbox{g}|$
does not show any measurable dependence on the P\'eclet number
\cite{CG77,PGM82,KRS91,MW98,TW94}. The corresponding pdf ${\cal
P}(\partial_{\|}\theta)$ is strongly asymmetric with a sharp maximum
located at $-g$, signaling the mean gradient expulsion out of the
plateaux regions. The same general picture is also valid for passive
scalar fields advected by synthetic flows.

The fact that the memory of the large-scale injection conditions is
not lost going toward small scales makes it necessary to have a better
understanding of scalar turbulence universality. To investigate the
degree and the type of dependence of scalar turbulence on the
injection mechanisms we have performed two series of numerical
experiments.  The advecting velocity field in both series is the
two-dimensional NS velocity discussed in Section~\ref{sec:ns}. The
major difference is in the injection mechanisms. As discussed in
Section~\ref{basic}, the first of them is isotropic (random forcing)
and the second is anisotropic (by a mean gradients $\bbox{g}$).  The
comparison between the two sets of results will permit a quantitative
analysis of universality.
\subsection{Anisotropy}
\label{pers_anis}

For the anisotropic injection by a mean gradient, the scalar structure
functions $S_{n}(\bbox{r})$ depend on both $r=|\bbox{r}|$ and
$\phi$, the angle between $\bbox{r}$ and $\bbox{g}$. The latter
dependence is best dealt with by expanding $S_{n}(r,\phi)$ on a proper
set of orthonormal functions $U^{(\bbox{l})}(\phi) $\,:
\begin{equation}
S_n(r,\phi)= \sum_{l} S^{(l)}_n(r) U^{(l)}(\phi).
\label{gener}
\end{equation}

The most natural basis is made of the set $\cos(l\,\phi)$, $\sin
(l\,\phi)$.  This is the two-dimensional version of the SO(d)
representation group exploited in Refs.~\cite{Arad98,Arad99} to
analyze experimental and numerical turbulence data. The sine functions
are in fact absent in the expansion (\ref{gener}), since the scalar
field is statistically invariant under the transformation $\phi\mapsto
-\phi$.  Furthermore, as a result of their opposite symmetries with
respect to the transformation $\bbox{r}\mapsto -\bbox{r}$
($\phi\mapsto \phi+\pi$), even and odd orders are expanded as
\begin{equation}
\label{decomp_even}
S_{2n}(\bbox{r})=\sum_{l=0}^n\,S_{2n}^{(2l)}\,(r)\,\cos\,(2\,l\phi)\,,
\end{equation}
\begin{equation}
\label{decomp_odd}
S_{2n+1}(\bbox{r})=\sum_{l=0}^n\,S_{2n+1}^{(2l+1)}\,(r)\,\cos\,[(2l+1)\phi]\,.
\end{equation}
Scaling behaviors $S_n^{(l)}(r)\propto r^{\zeta_n^{(l)}}$ are expected
in the inertial range. The n-th order structure function is thus made
of a power law superposition. It is expected that the various
contributions are ordered according to their anisotropic degree,
i.e. $\zeta^{(l)}_n$ increases with $l$ for a fixed $n$ \cite{Arad99}.

To extract the leading contribution to the scaling of
$S_{n}(\bbox{r})$, it is sufficient to project on the lowest-order
components as
\begin{equation}
S_{2n}^{(0)}(r)=\frac{1}{2\pi}\,\int_0^{2\pi} S_{2n}(r,\phi)\, d\phi, 
\label{pari}
\end{equation}
\begin{equation}
S_{2n+1}^{(1)}(r)=\frac{1}{2\pi}\,\int_0^{2\pi} 
S_{2n+1}(r,\phi)\,\cos\phi\, d\phi. 
\label{dispari}
\end{equation}
The scaling behaviors presented hereafter have been obtained by
performing such angular averages over a set of 80 snapshots equally
spaced by about one turn-over time.  We have also verified that the
contributions arising from higher anisotropic sectors are subdominant
as expected, i.e. that the exponents associated to
\begin{equation}
S_{n}^{(l)}(r)=\frac{1}{2\pi}\,\int_0^{2\pi} 
S_{n}(r,\phi)\,\cos(l \phi)\, d\phi\qquad l \geq 2\,, 
\label{sublead}
\end{equation}
are larger than those for $l=0$ or $l=1$. A precise measurement of the
subdominant anisotropic scaling exponents is rather delicate due to
the cancellations needed for their statistical convergence and we
shall not dwell on this.

A remark on the projection procedure is in order. In numerical
simulations, it is easy to perform angular averages such as
(\ref{pari}) or (\ref{dispari}). This is not generally the case for
laboratory experiments where measuring the data for all possible
directions $\phi$ might be problematic. It is then useful to note that
the results obtained by using the projections (\ref{pari}) or
(\ref{dispari}) essentially coincide with those where only the first
subleading contributions are eliminated. This filtering is much more
economic as it can be performed by a single measurement along
$\phi=\pi/4$ for even orders and $\phi=\pi/6$ for odd-order structure
functions.

\narrowtext
\begin{figure}
\epsfxsize=7.6truecm
\epsfbox{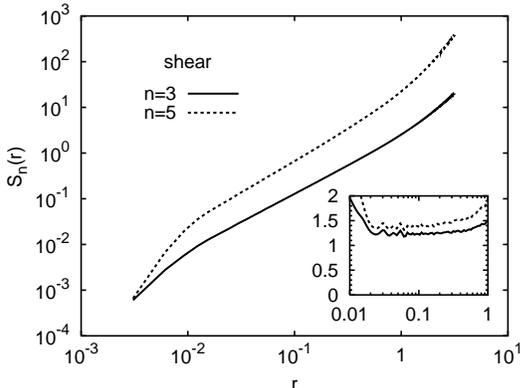}
\caption{The 3rd and the 5th-order parallel structure functions for
the injection by a mean gradient. In the inset, local scaling
exponents $dS_n(r)/d\log r$. The measured exponents are
$\zeta_3=1.25\pm 0.04$, $\zeta_5=1.38\pm 0.07$, with error bars
estimated from rms fluctuations of local scaling exponents.}
\label{oddstru}
\end{figure}
Let us now present the results concerning the anisotropy decay rate,
measured by odd-order structure functions.
In Fig.~\ref{oddstru}, we
show the $3$-rd and the $5$-th order structure functions.  Accordingly,
the skewness and hyperskewness coefficients of scalar differences
scale as
\begin{equation}
S_3/S_2^{3/2} \sim r^{0.25}, \qquad S_5/S_2^{5/2} \sim r^{-0.2},
\label{trecinque}
\end{equation}
the second-order scaling exponent being compatible, within the error
bars, with the dimensional value $\simeq 2/3$ (see
Fig.~\ref{evenstru}). Both behaviors (\ref{trecinque}) violate the
dimensional prediction (\ref{ske}). Furthermore, whilst the skewness
is decaying, even though much more slowly than the expected
$r^{2/3}$, the hyperskewness grows at small scales. The effect of
memory of the large-scale injection conditions observed in laboratory
experiments is thus dramatically present also in our numerical
simulations. The fact that clean scaling behaviors and exponents are
measured here gives a very strong indication in favor of the fact
that we are not dealing with a finite P\'eclet number effect. The most
natural conclusion stemming from experiments, the previous and the
present numerical simulations is that isotropy is not restored in the
full classical sense, no matter how large is the P\'eclet number.

\begin{figure}
\epsfxsize=7.6truecm
\epsfbox{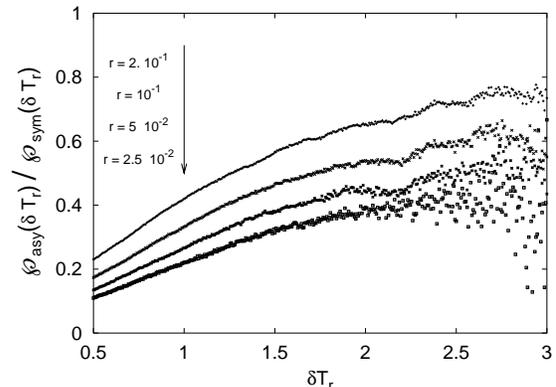}
\caption{Ratio between the anti-symmetric and the symmetric part of 
scalar increment pdf's ${\cal P}(\delta_r T)$ for four separations in 
the inertial range.}
\label{sym_asym}
\end{figure}

It is worth discussing in some more detail the relation between this
last conclusion and the observation previously made on the
subdominance of higher-order anisotropic contributions to structure
functions. The problem comes of course from the presence of
intermittency\,: contrary to the classical theory, the pdf of scalar
differences ${\cal P}(\delta_r T)$ for various separations $r$ cannot
be collapsed one onto another by a simple rescaling. Their change of
shape with the separation $r$ makes it crucial to specify how the
anisotropy degree is quantified and the amplitude of the fluctuations
sampled in the pdf's at various $r$'s. The ratio $S_5/S_4^{5/4}$ is
for example decreasing with $r$, contrary to the hyperskewness in
(\ref{trecinque}). The blow-up of the latter is thus a joined effect
of anisotropy and intermittency.  The point to be remarked is that
odd-order structure functions are all scaling with exponents higher
than those of structure functions of the same order but calculated
with the absolute values, e.g. $\int |\delta_r T|^3\,{\cal P}(\delta_r
T)\,d \delta_r T$. In other words, when fluctuations of similar
amplitude are compared, anisotropic contributions are subdominant with
respect to the isotropic ones.  The same point is stressed in Fig.~\ref{sym_asym},
where it is shown that the ratio between the anti-symmetric and the
symmetric part of scalar increment pdf's decreases with
$r$. Notwithstanding the conclusion on the absence of isotropy
restoration in the classical sense, it is thus important to realize
that anisotropic fluctuations become more and more subdominant as
their anisotropy degree increases.
\narrowtext
\subsection{Universality}

What is the degree of universality of scalar turbulence with respect
to the injection mechanisms? Is the persistence of anisotropy just
discussed signalling that the scalar statistics is totally imprinted
by the non-universal large-scale injection conditions even at the
smallest scales? To answer these questions it is convenient to compare
observables, such as even-order structure functions or pdf's
themselves, that are meaningful and non trivial for both types of
injection.

\begin{figure}
\epsfxsize=8truecm
\epsfbox{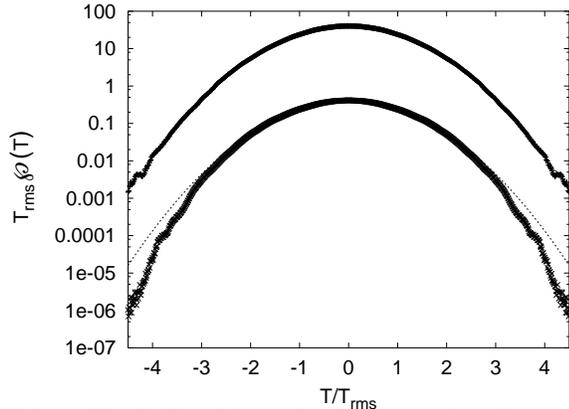}
\caption{The Pdf's of the scalar field normalized by its standard
deviation for the two cases of random forcing (upper curve,
multiplied by a factor $10^2$ for plotting purposes) 
and injection by a mean gradient (lower curve). 
Note the presence of subgaussian tails for the latter
(Gaussian densities are shown for comparison as dotted lines).}
\label{f4}
\end{figure}

Let us first start from the scalar field pdf. Even though this is not
a small-scale quantity, it is interesting that the two curves are
different and the one for the mean-gradient injection is subgaussian,
see Fig.~\ref{f4}. Note that this is not in contradiction with the
exponential behavior predicted in \cite{SSP,SS94,Cetal95}. The latter
work considers indeed the situation where the velocity correlation
length is much smaller than the typical length of variation of the
large-scale mean scalar profile. In our case, these two lengths are of
the order of the velocity friction length scale, defined in
Section~\ref{models}, and of the box size, respectively. They are thus
comparable. Similar subgaussian behaviors were also observed in
\cite{JW92}.
\narrowtext

\begin{figure}
\epsfxsize=8truecm
\epsfbox{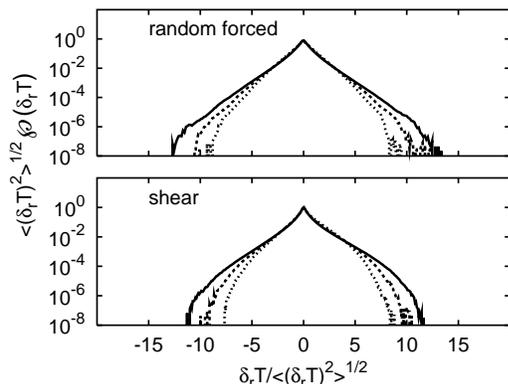}
\caption{Pdf's of scalar increments normalized by their standard
deviations for three separations $r=2.5\times\,10^{-2}$,
$5\times\,10^{-2}$, $10^{-1}$ in the inertial range.}
\label{isanispdf}
\end{figure}

\begin{figure}
\epsfxsize=8truecm
\epsfbox{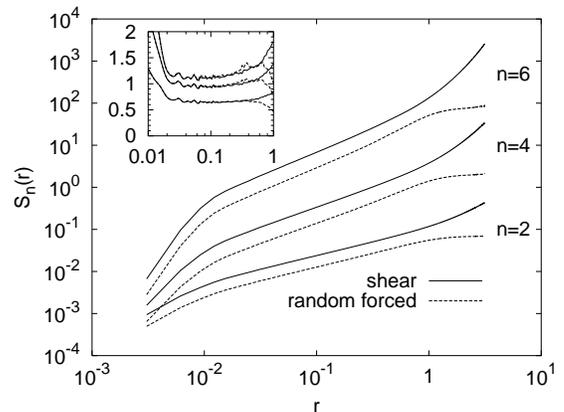}
\caption{Low-order even structure functions. Local scaling exponents
are shown in the inset. The measured exponents are
$\zeta_2=0.66\pm 0.03$, $\zeta_4=0.95\pm 0.04$ and $\zeta_6=1.11 \pm
0.04$.}
\label{evenstru}
\end{figure}

Let us move then to the small-scale statistics by considering scalar
increments. In Fig.~\ref{isanispdf}, we present the probability
density functions for the two injection mechanisms\,: it is clear that
they do not have the same shape and the same conclusion holds if we
take only the symmetric part of the pdf's. A particular consequence is
that structure functions cannot have a universal expression, as
postulated in the classical theory. It is however directly checked
from Fig.~\ref{evenstru} that the scaling exponents are the same for
the two types of injection. This implies that the pdf's, although
having different shapes, rescale with $r$ in the same way. 
\narrowtext
The scaling
exponents being universal, the dependence on the large-scale injection
conditions shows up in the structure function dominant expression
$S_n(r)=C_n
\epsilon_{\theta}^{n/2}\epsilon_{v}^{-n/6}r^{n/3}(L/r)^{n/3-\zeta_n}$
via the non-dimensional constants $C_n$.
The resulting picture of universality is the following\,: structure
function scaling exponents are independent of the injection details
and thus universal\,; scalar and scalar increment pdf's, and
non-dimensional constants in the structure functions are not. A
particular instance is the one of odd-order structure functions, where
the coefficients $C_{2n+1}$ vanish or not, depending on the symmetries
of the injection mechanisms. This form of universality is weaker than
the one in the original classical KOC theory and agrees with the
remark made by L.D. Landau about the universality in Kolmogorov 1941
theory for fully developed turbulence. Remark finally that this
universality framework is the same as the one arising from the zero
mechanism in the Kraichnan passive scalar model \cite{GK95,CFKL95,SS95}.

\narrowtext

\section{Saturation of intermittency}
\label{int-satu}

The cliffs observed in scalar fields are strikingly suggestive of
quasi-discontinuities. When smaller and smaller molecular
diffusivities are considered, the minimal width of the fronts shrinks
with the dissipation scale, with their maximum amplitude remaining
comparable to the scalar rms value.  Simple phenomenology suggests
that the presence of such structures, corresponding to a local
H\"older exponent equal to zero, might induce a vanishing slope in the
structure function scaling exponent curve. The fronts being the
strongest possible events, this behavior should take place for large
enough orders, whence the possible saturation $\zeta_n\to {\rm const}$
for high $n$'s.

The first and most natural way to investigate the saturation is to
look directly at the scalar structure function scaling exponents.  An
alternative procedure consists in looking, for a fixed $\delta_r T$,
at how the pdf ${\cal P}(\delta_rT)$ of the scalar differences varies
with the separation $r$. The latter is statistically more reliable
than the former as less sensitive to the extreme tails of the pdf.
Saturation is equivalent to the pdf taking the form
\begin{equation} {\cal P}(\delta_r T)= r^{\zeta_{\infty}}{\cal Q}\left
(\frac{\delta_r T}{T_{rms}}\right )\frac{1}{T_{rms}}, \label{pdf}
\end{equation} for values of $\delta_r T\gtrsim T_{rms} \equiv \left
(\langle T^2 \rangle - \langle T\rangle^2\right )^{1/2}$.  The tails
of the (non-universal) ${\cal Q}$ function are bounded by the
single-point scalar pdf, as it follows from the simple inequality
(see, e.g., Ref.~\cite{NWLMF97})
\begin{equation} Prob\,\{|\delta_r T|>
\tau\}\leq 2\, Prob\,\{|T|>\tau/2\}. \end{equation} 

Both procedures have been used for the case of the Navier-Stokes
advecting velocity field, where we could efficiently use spectral
numerical methods. They present the disadvantage of being sensitive to
finite-size effects but the advantage of giving access to the whole
field and thus to the pdf's. For the Kraichnan model, spectral methods
are not very efficient and it is more convenient to use Lagrangian
techniques (see Ref.~\cite{FMV98,FMNV99} for a detailed
discussion). The latter permit very precise measurements of the
scaling exponents at the price of not giving access to the pdf's.

\subsection{Saturation with the Navier--Stokes advecting flow}
\label{res-ns2d}

\begin{figure} 
\epsfxsize=7.6truecm 
\epsfbox{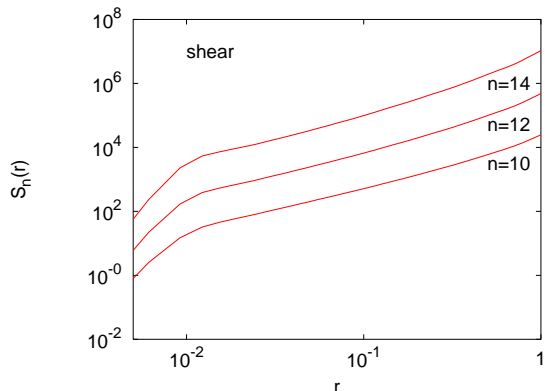}
\caption{Scalar structure functions of orders $10, 12$ and $14$
for the anisotropic injection mechanism}
\label{s10-12-14} \end{figure}

\begin{figure} 
\epsfxsize=7.6truecm 
\epsfbox{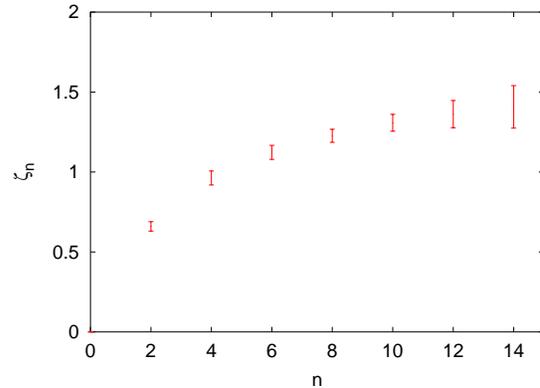}
\caption{Measured scaling exponent $\zeta_n$ for the Navier--Stokes
advection. Error bars are estimated by
the rms fluctuations of local scaling exponents.}
\label{zn-vs-n_ns2d} \end{figure}

\begin{figure} 
\epsfxsize=7.6truecm 
\epsfbox{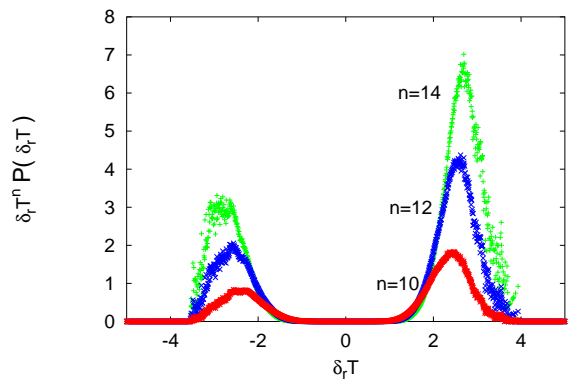}
\caption{Bulk contribution to the $n$-th moment,
$\delta_r T^n \cal P(\delta_r T)$, for $n=10,12,14$.
Here $r$ is at the lower end of the inertial range.}
\label{test}
\end{figure}

\begin{figure} 
\epsfxsize=7.6truecm 
\epsfbox{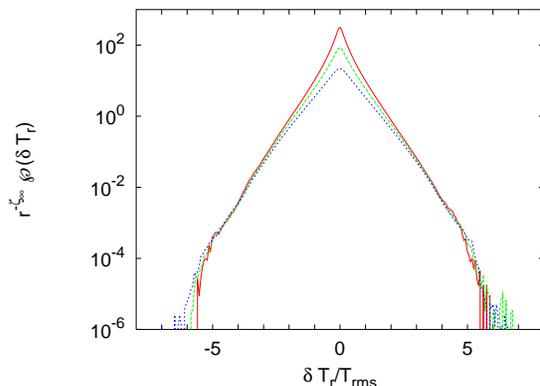}
\caption{The pdf's ${\cal P}(\delta_r T)$ , for the random forced injection,
 for three values of $r$
inside the inertial range of scales, multiplied by the factor
$r^{-\zeta_\infty}$.  The collapse of the curves indicates the
presence of saturation and also gives the unknown function ${\cal Q}
(\delta_r T/T_{rms})$ in Eq.~(\protect\ref{pdf}).}  
\label{collasso}
\end{figure}

\begin{figure} 
\epsfxsize=7.6truecm 
\epsfbox{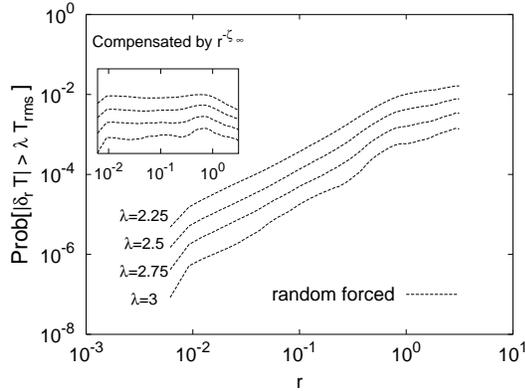}
\caption{Probabilities $\int_{\lambda T_{rms}}^{\infty}{\cal P}$ {\it
vs} $r$ of having $\delta_r T>\lambda T_{rms}$ for various $\lambda$.
In the inside zoom, such probabilities are compensated by the factor
$r^{-\zeta_\infty}$.}
\label{cumul-pdf}
\end{figure}

\begin{figure} 
\epsfxsize=7.6truecm 
\epsfbox{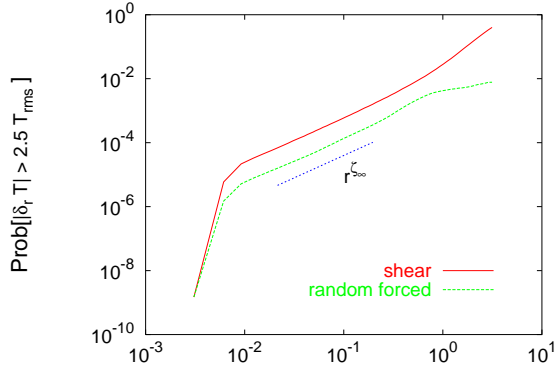}
\caption{Probabilities $\int_{2.5\,T_{rms}}^{\infty}{\cal P}$ {\it vs}
$r$ of having $\delta_r T>2.5 T_{rms}$ for both the isotropic and
non-isotropic injection mechanisms. The parallelism of the curves
reflects the universality of the saturation exponent
$\zeta_{\infty}$.}
\label{univ}
\end{figure}

The behaviors of the structure functions $S_{10}$, $S_{12}$ and
$S_{14}$ in the presence of a large-scale scalar gradient are shown
in Fig.~\ref{s10-12-14}.
The scaling exponents $\zeta_n$ {\it vs} the order $n$ (up to the
order $14$) are reported in Fig.~\ref{zn-vs-n_ns2d}. The curve is
compatible with a saturation exponent $\zeta_{\infty}\sim 1.4$.  The
convergence of the moments was inspected by the usual test of checking
whether $(\delta_r T)^{14} {\cal P}(\delta_r T)$ decays before the pdf
${\cal P}$ becomes noisy.  The bulk contribution to the moments
$10,12$ and $14$ is shown in
Fig.~\ref{test}.
As for the second procedure on the pdf's, following (\ref{pdf}) we
plot in Fig.~\ref{collasso} the curves $r^{-\zeta_\infty}{\cal
P}(\delta_r T)$. Their collapse is a signature of the saturation and
gives the unknown function ${\cal Q}$.
\narrowtext 
\narrowtext 
\narrowtext 
\narrowtext 
In Fig.~\ref{cumul-pdf} we show
the probabilities $\int_{\lambda\;T_{rms}}^{\infty}{\cal P}$ {\it vs}
$r$ of having $\delta_r T >\lambda T_{rms}$ for various values of
$\lambda$. The parallelism of the curves is again the footprint of
saturation. Explicit evidence for the universality of $\zeta_{\infty}$
is provided in Fig.~\ref{univ}, where the probabilities
$\int_{2.5\;T_{rms}}^{\infty}{\cal P}$ of having $\delta_r T
>2.5\,T_{rms}$ are shown both for the isotropic and the non-isotropic
injection mechanisms.
\narrowtext 
\narrowtext 

Both procedures give therefore a strong evidence in favor of
saturation for the Navier-Stokes advecting velocity field. 

\subsection{The case of the Kraichnan model}
\label{sec:sat-rhk}

The advecting velocity field in the Kraichnan model \cite{K94} is
assumed to be Gaussian, homogeneous, isotropic, of zero mean and
correlation function
\begin{eqnarray}
\label{correlation}
\langle \left( v(\bbox{r},t)-v(\bbox{0},t)\right)_{\alpha} \left(
v(\bbox{r},t)-v(\bbox{0},t)\right)_{\beta}\rangle= \\
D\,r^{\xi}\left[(d+\xi-1)\delta_{\alpha\beta}-\xi {r_{\alpha}
r_{\beta}\over r^2}\right],
\end{eqnarray}
where $d$ is the dimension of space. The assumption of
$\delta$-correlation in time is of course far from the reality, but it
has the remarkable feature of leading to closed equations for
equal-time correlation functions
$C_{n}\equiv\left\langle\theta(\bbox{r}_1,t)\ldots\theta
(\bbox{r}_n,t)\right\rangle$ of any order $n$ (see, e.g.,
Refs.~\cite{K68,SS94}).  This has permitted to rely the emergence of
anomalous scaling and intermittency to the so-called zero modes, i.e.
functions annihilated by the inertial operators appearing in the
aforementioned closed equations ~\cite{GK95,CFKL95,SS95}. Anomalous
zero modes can be calculated nonperturbatively for the case of a
passively advected magnetic field \cite{MV} and for shell models {\it
\`a la} Kraichnan \cite{BBW97}. 
For the case of the passive scalar, the
expression for the second order correlation function $C_{2}$ is
known exactly \cite{K68} and the inertial-range scaling behavior for
the second-order structure function reads:
\begin{equation}
S_2(r;L)=\frac{2\chi(0)}{\zeta_2(d-1)d D}\, r^{\zeta_2}\qquad \zeta_2=2-\xi,
\end{equation}
where $\chi(0)$, defined in (\ref{fcorr}), is the energy injection
rate.  The exponent $\zeta_2$ coincides with the predictions based on
dimensional arguments and $\xi=4/3$ is the velocity exponent
corresponding to the KOC scaling of the scalar field.  Exact solutions
are not available for higher orders and analytical predictions are
perturbative in three different limits of the model\,: rough velocity
fields $\xi\ll 1$ (Refs.~\cite{GK95,P96,AAV98})\,; large space
dimensionalities $d\gg 1$ (Ref.~\cite{CFKL95}) and almost smooth
velocity fields $\xi\to 2$ (Ref.~\cite{SS95}). In all of them, the
order $n$ of the structure functions enters into the correction to the
dimensional scaling together with the small parameter of the
expansion. This means that, for any non-vanishing value of the
expansion parameter, the predictions for $\zeta_n$ will break down at
large enough $n$ and the very strong events associated to the extreme
tails of the pdf's cannot be captured by such procedures.  To deal
with the rare fluctuations responsible for the asymptotic behavior of
the exponents $\zeta_n$ at large $n$'s, instantonic approaches have
been borrowed from field theory \cite{Li76,ZJ}.  In the limit of
very high $d$'s, the problem could be solved \cite{BL98}, with
saturation taking place for $n\ge n^*$ (the latter diverging 
as $d\to\infty$).  In the three-dimensional case, saturation has been
phenomenologically suggested in Ref.~\cite{Y97} and inferred from an
instantonic bound in Ref.~\cite{C97}.

Since numerical simulations are {\it a fortiori} limited to the
measure of finite-order structure functions, it is important to
choose the most convenient conditions to observe saturation. In
particular, we expect that, for a fixed dimensionality of space, the
orders where saturation takes place reduce with $\zeta_2=2-\xi$. This
is due to the mechanism of cliff steepening. The latter is clearly
favored by the presence of large-scale velocity components, inducing
a coherent shear effect, and obstacled by small-scale velocity
components, leading to effective diffusion effects. Since the
amplitude of the former increases as $2-\xi$ reduces, we expect that
saturation is favored by choosing $\xi$ close to two.  Approaching too
closely the Batchelor limit $\xi=2$ is however problematic from a
numerical point of view, due to the presence of strong nonlocal
effects. The chosen trade-off is to consider the three cases
$\xi=1.875$, $\xi=1.84$, $\xi=1.75$, in 3D.  The exponents being
universal, it is more convenient to consider the isotropic injection
mechanism (\ref{fcorr}) and we shall look at the fourth and the
sixth-order structure functions.
\narrowtext \begin{figure} 
\epsfxsize=6.6truecm 
\epsfbox{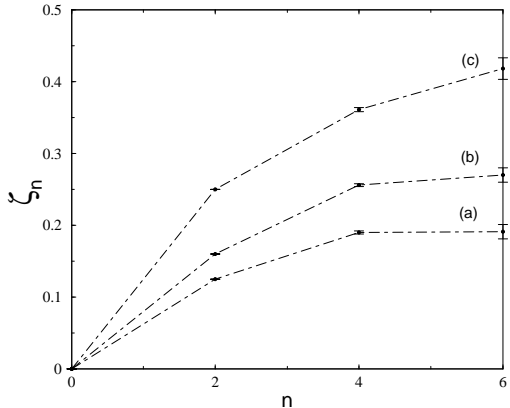}
\caption{The measured scaling exponent $\zeta_n$'s joined by straight-lines.
(a) $\xi=1.875$; (b) $\xi=1.84$ and (c) $\xi=1.75$.}
\label{zeta_vs_n}
\end{figure}

A detailed description of the Lagrangian method can be found in
Refs.~\cite{FMV98,FMNV99} and will not be reported here. We only
remark the fact that it naturally allows to measure the scaling of the
structure functions $S_{2n}(r;L)$ {\it vs} the integral scale $L$ of
the forcing. Physically, this means that the injection rate of the
passive scalar variance (which equals its dissipation rate) and the
separation $r$ are kept fixed, while the integral scale $L$ is varied.
Anomalies, that is deviations from dimensional scaling exponents, are
therefore measured directly (see (\ref{evenmom})) through the scaling
dependence on $L$ of the structure functions.

\begin{figure} 
\epsfxsize=7.2truecm 
\centerline{\epsfbox{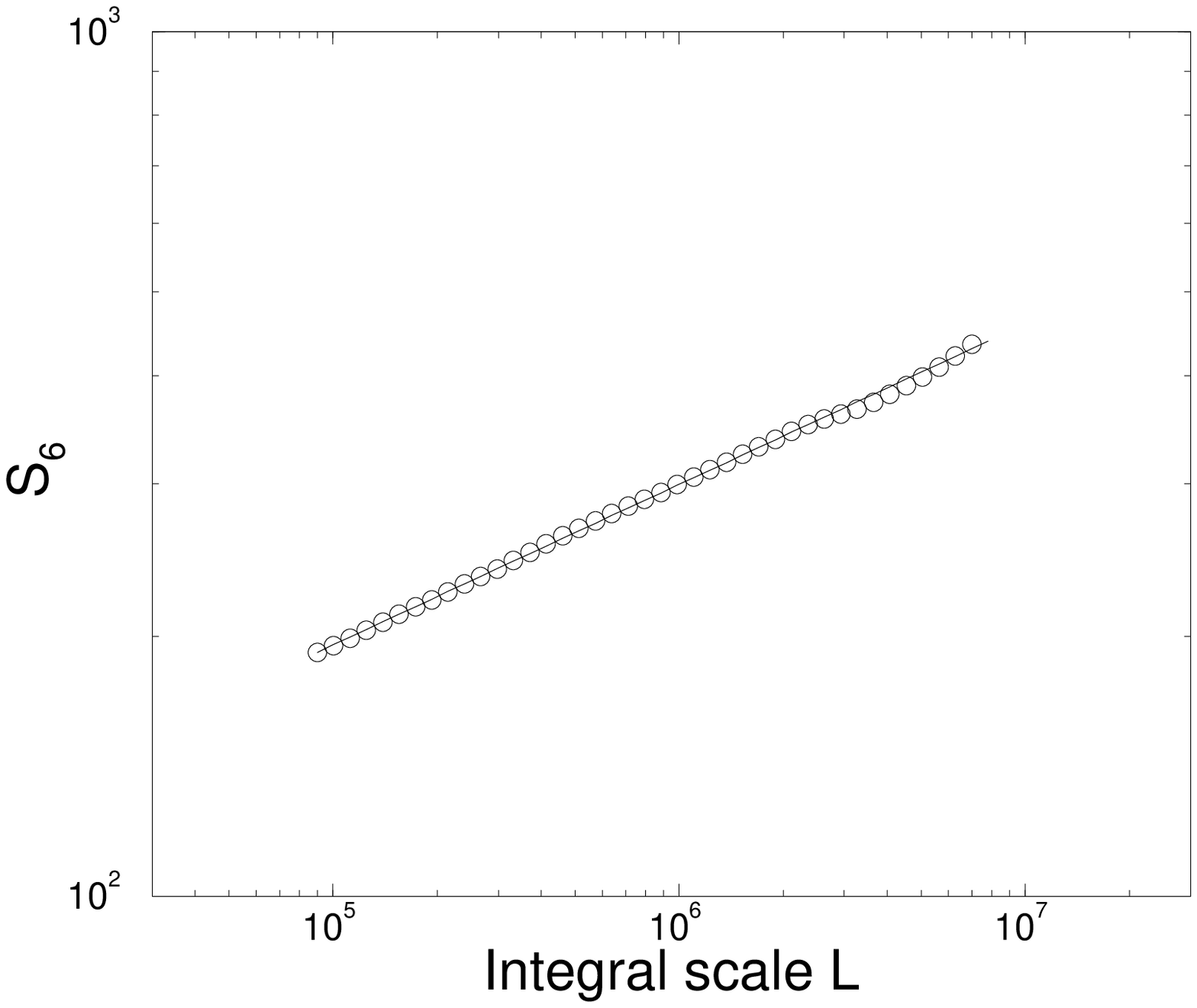}}
\epsfxsize=7.2truecm 
\centerline{\epsfbox{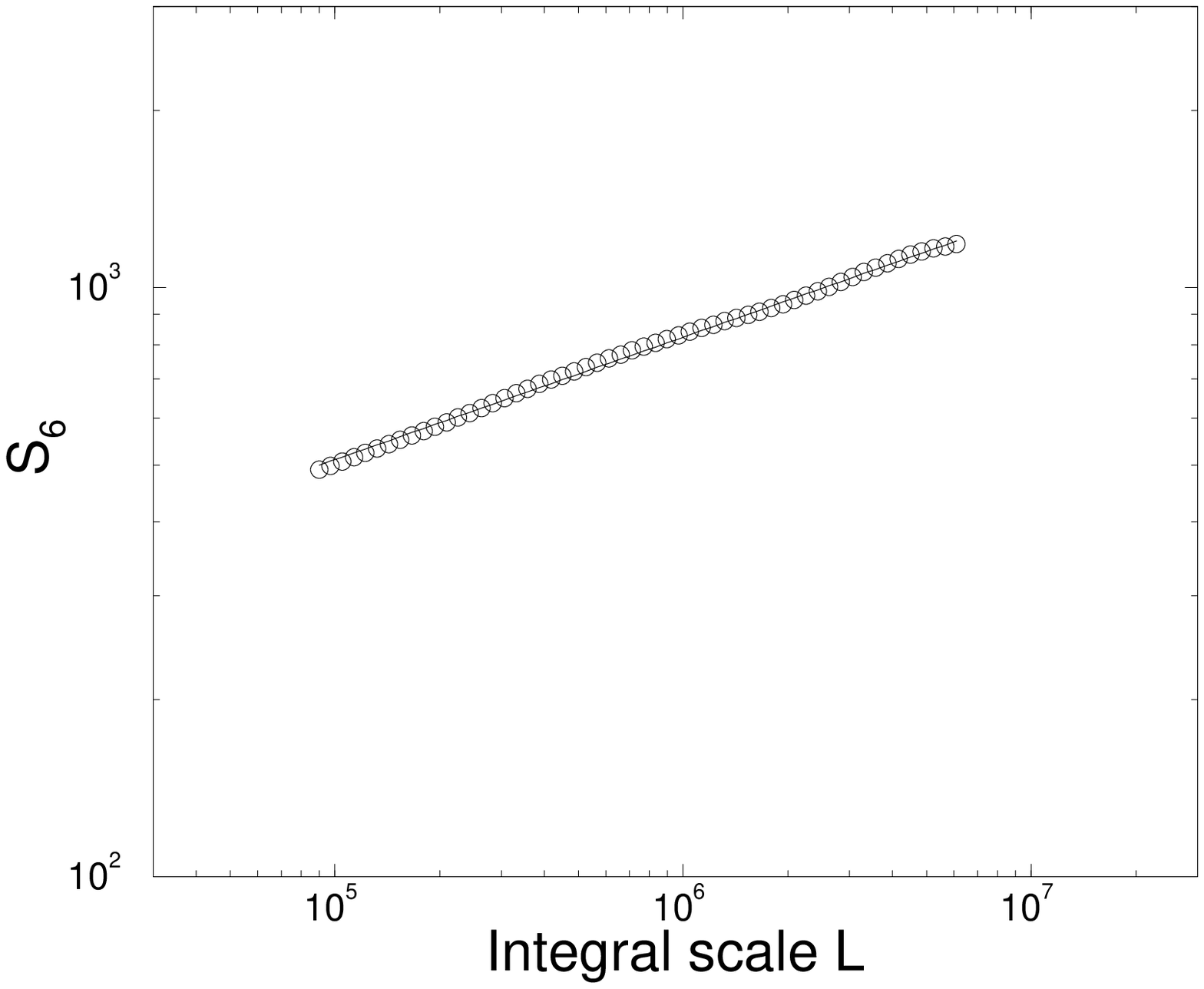}}
\epsfxsize=7.2truecm 
\centerline{\epsfbox{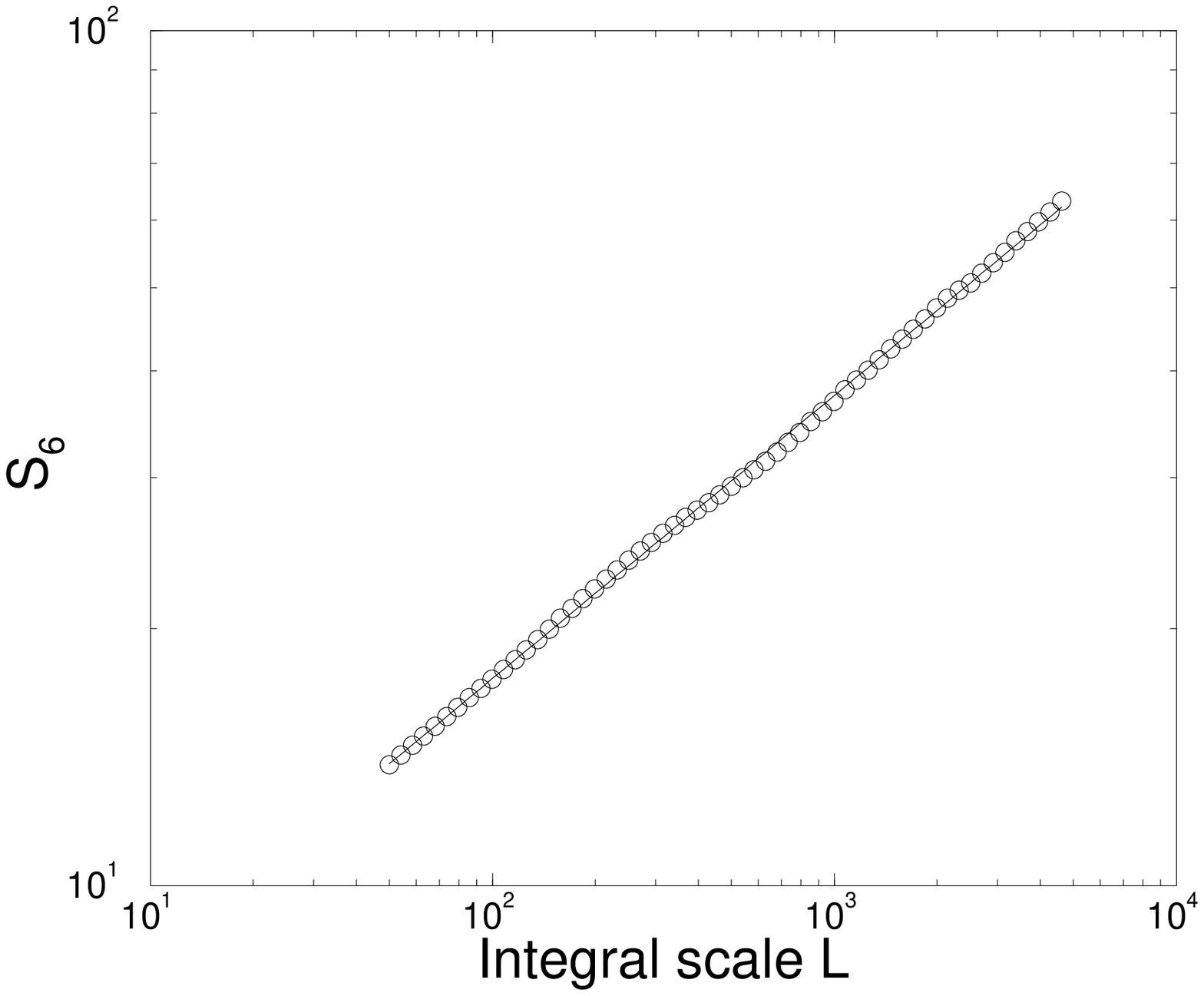}}
\caption{3-d sixth order structure function $S_6$ {\it vs} $L$. 
Top: $\xi=1.875$. Parameters: $r=0.027$, $\kappa=5\times 10^{-12}$
and number of realizations $\sim 21\times 10^{6}$. 
The best-fit (solid) line has the slope 
$\rho_{6}=3\/\zeta_2- 0.19=0.185\;(\pm 0.01)$.
Middle: $\xi=1.84$. Parameters: $r=0.54$, $\kappa=5\times 10^{-12}$
and number of realizations $\sim 23\times 10^{6}$. 
The best-fit (solid) line has the slope $\rho_{6}=3\/\zeta_2-0.27
=0.21\;(\pm 0.01$).
Bottom: $\xi=1.75$.  
Parameters: separation $r=0.032$, diffusivity 
$\kappa=1\times 10^{-9}$
and number of realizations $\sim 27\times 10^{6}$.
 The best-fit (solid) line has the slope 
$\rho_{6}=3\/\zeta_2-0.42=0.33\;(\pm 0.015)$.}
\label{sei_xi}
\end{figure}

\begin{figure} 
\epsfxsize=7.2truecm 
\epsfbox{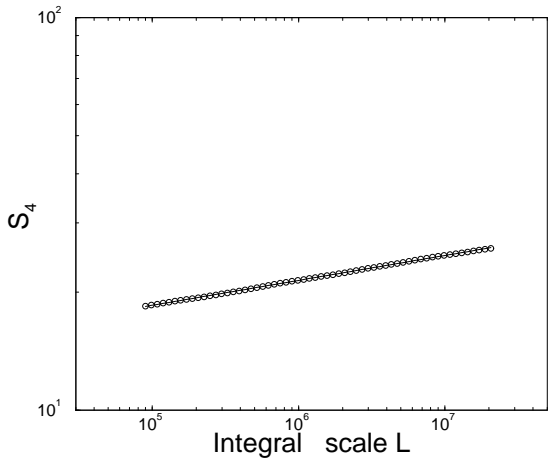}
\caption{3-d fourth order structure function $S_4$ {\it vs} $L$
for $\xi=1.875$. Parameters: $r=0.027$, $\kappa=5\times 10^{-12}$
and number of realizations $\sim 5\times 10^{6}$. 
The best-fit (solid) line has the slope 
$\rho_{4}=2\/\zeta_2- 0.064=0.186\;
(\pm 0.002)$.}
\label{quattro_xi1.875}
\end{figure}

The measured $\zeta_n$'s are shown in Fig.~\ref{zeta_vs_n} for:
(a) $\xi=1.875$, (b) $\xi=1.84$ and (c) $\xi=1.75$. 
(The error bars are estimated by
the rms fluctuations of local scaling exponents over intervals 
of constant length $\log_{10} 5$ in log-space.)
The corresponding
behaviors of $S_6$ 
{\it vs} the integral scale $L$ are shown
in Fig.~\ref{sei_xi}  for 
$\xi=1.875$, $\xi=1.84$ and $\xi=1.75$, respectively. 
The structure function $S_4$ for $\xi=1.875$ is shown in 
Fig.~\ref{quattro_xi1.875}.
Similar high quality scaling laws have been obtained for $\xi=1.84$
and $\xi=1.75$.
Some remarks about Fig.~\ref{zeta_vs_n} are worth.  It follows from
H\"older inequalities that the $\zeta_n$ curve for $n>6$ must lie
below the straight line joining $\zeta_4$ and $\zeta_6$.  It seems
unlikely that $\zeta_n$ can be a decreasing function of $n$. Indeed,
an adaptation of the argument presented in Ref.~\cite{UF95}, indicates
that a decreasing $\zeta_n$ would entail arbitrarily large temperature
differences at very small scales, something unphysical, given the
maximum principle for the advection diffusion equation. Note that {\em
stricto sensu\/} this argument applies only to the unforced equation,
but the scalar scaling exponents are expected to be the same in the
forced and in the unforced case (in its quasi-stationary
phase). Hence, curve (a) in Fig.~\ref{zeta_vs_n} and H\"older
inequalities strongly constrain the behavior at higher $n$'s. 
\narrowtext 
%
%
%
\narrowtext 
Of course, the presence of error bars makes it impossible, with a finite
number of measured $\zeta_n$'s, to state rigorously that the curve (a)
tends to a constant. It is nevertheless quite clear that the latter
provides strong evidence in favor of saturation of the scaling
exponents of the structure functions.  From curves (b) and (c) we can
also notice that the dependence on $\xi$ of the order $n$ where
saturation takes place is indeed increasing with $\zeta_2=2-\xi$, in
qualitative agreement with our previous simple physical arguments and
the result in Ref.~\cite{BL98}.

\section{Fronts}

\begin{figure} 
\epsfxsize=8truecm 
\epsfbox{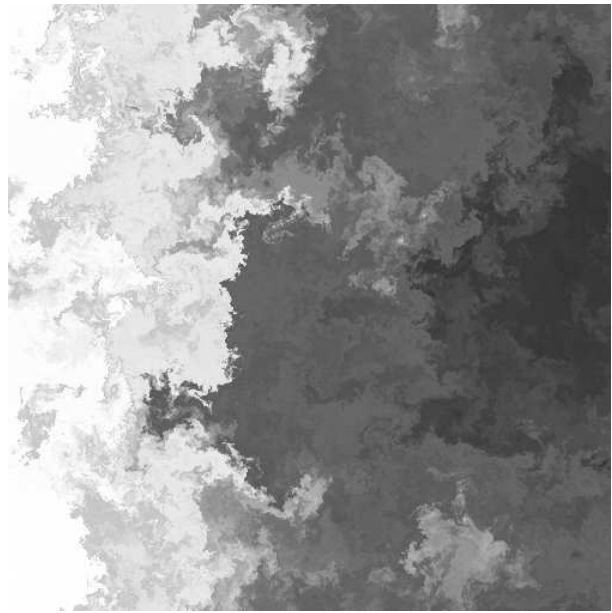}
\vskip 5pt
\caption{A snapshot of the scalar field, $T$, 
for the injection by a mean gradient directed from right to left.  
Colors are coded according to the intensity of the field: white 
corresponds to large positive values, black to large negative values.}
\label{r-and-c} \end{figure}

The presence of fronts in the scalar field 
(See Fig.~\ref{r-and-c}) is a major characteristic of scalar turbulence.
How is it related to intermittency saturation ?
As we have shown in Fig.~\ref{cumul-pdf}, the probability to have a 
fluctuation which is $O(T_{rms})$ across a separation $r$ goes as
a power law $r^{\zeta_{\infty}}$. This result admits two different
interpretations which are associated to very different physical pictures. 

\begin{figure} 
\epsfxsize=10truecm 
\epsfbox{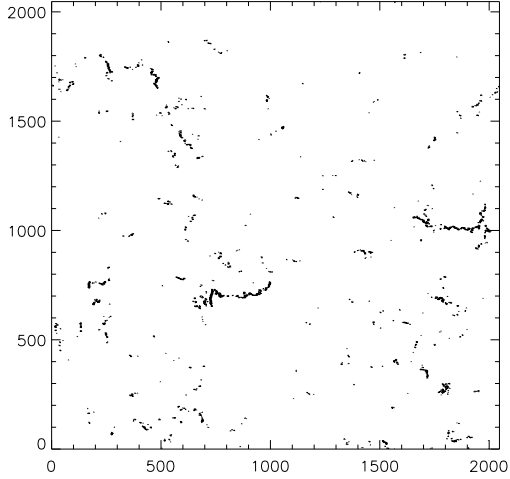}
\caption{The union of the sets where mature fronts are located. The
snapshots over which the union is performed are $80$, regularly spaced
by one eddy turnover time.}
\label{f9} \end{figure}

As for the first scenario, assume that the most important contribution
to excursion of $O(T_{rms})$ comes from jumps of the scalar field
occurring across a very small lengthscale (comparable to the diffusive
scale $\eta \sim (\kappa^3/\epsilon_v)^{1/4}$).  We shall call these
jumps {\em mature fronts}, to indicate that the process of gradient
steepening is brought to its utter development.  Then, the probability
of observing large scalar fluctuations at scales larger than $r$ is
determined by geometrical considerations. Suppose, for instance, that
the fronts have an extension $O(L)$ in the direction transverse to the
jump, that is the gradients $O(T_{rms}/\eta)$ are occurring on a set
of points which form a line. The probability of intercepting a front
across a distance $r$ is then proportional to $r$ itself.  In this
case, the saturation exponent would be equal to unity. Since the measured
exponent for the probability is $\zeta_{\infty} \simeq 1.4$ we infer
that mature fronts live on a set of dimension $D_F\simeq 0.6$. A
dimension smaller than one means that the largest gradients occupy a
fraction of space smaller than that occupied by a smooth line, thus
mature fronts appear as a collection of short, broken lines where a
steep gradient takes place (see Fig.~\ref{f9}).  What is peculiar of
this scenario is that, as the P\'eclet number goes to infinity, the
thickness of the fronts decreases, and larger gradients take place,
but the probability of finding a mature front remains constant.  This
physical picture has some analogies with that encountered in Burgers
turbulence\,: there, it is well-known \cite{AFLV92} that the presence
of isolated shocks (i.e.~mature cliffs of unit co-dimension),
connected by smooth ramps, is the cause of intermittency saturation.
In the context of this purely geometrical interpretation the time
variability does not play any role, which amounts to say that mature
fronts should be long-living objects. This remark leads us to consider
a different possibility.
\narrowtext \vskip -2pt
The second scenario is based on the idea that the process of front
formation starts from a $O(T_{rms})$ fluctuation of the scalar field
generated at large scales by injection mechanisms. This profile gets
steepened by the action of large-scale converging components of the
velocity field, until the stretching exerted by small-scale velocity
fluctuations overcomes the tendency of the front to steepen.  As a
result, the front survives only down to a scale $r$, larger than the
dissipative scale $\eta$. The probability that the steepening process
stops at a scale $r$ decreases with $r$ according to
$r^{\zeta_{\infty}}$. At variance with the former picture, here the
probability of observing a mature front vanishes as
$\eta^{\zeta_{\infty}}$, thus it is strictly zero for an infinite
P\'eclet number.

\begin{figure} 
\epsfxsize=7.6truecm 
\epsfbox{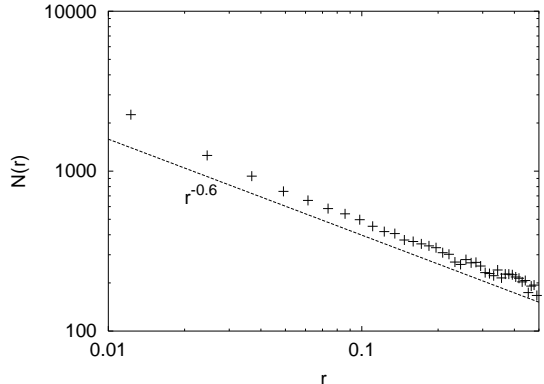}
\caption{The box-covering dimension of the mature cliffs set. 
The number of boxes of size $r$ necessary to cover the set
represented in Fig.~\ref{f9} decays as $N(r)\sim r^{-D_F}$ with
$D_F=0.6$.}
\label{f10} \end{figure}

It has to be remarked that in both cases, the dimension
of the set hosting mature fronts has  to be $D_F=2-\zeta_{\infty}$, but
the number of points which belong to this set is vanishing with 
increasing P\'eclet.
Thus, the observation of a dimension $D_F\simeq 0.6$ 
(see Fig.~\ref{f10} ) does not allow us
to discriminate between the first and the second picture.
\narrowtext 
\narrowtext 
In a nutshell, in the first scenario intense small-scale excursion of
scalar field are essentially generated by mature fronts, whereas in
the second one mature fronts play a statistically insignificant role.

\begin{figure} 
\epsfxsize=8truecm 
\epsfbox{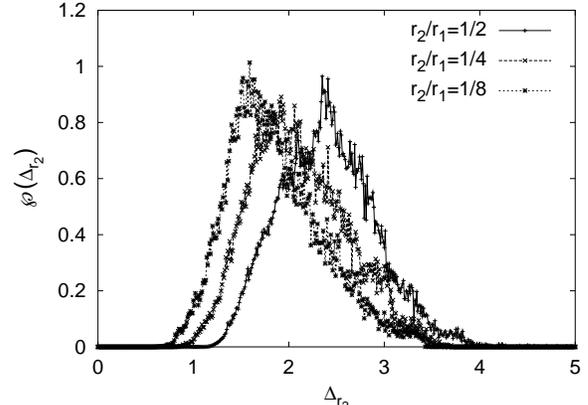}
\caption{The pdf of $\Delta_{r_2}$ -- maximal excursion across a
length $r_2$ -- under the condition that an excursion larger than $2.5
T_{rms}$ takes place at a scale $r_1$, for three different
$r_2/r_1$. All distances belong to the inertial range.}
\label{f6} \end{figure}

In order to quantify the weight of mature fronts in the build-up of
strong fluctuations and thus distinguish among the two options listed
above, we computed the following conditional probability. Given a jump
of height $|\delta_{r1} T| = O(T_{rms})$ across an interval $I_1$ of
length $r_1$, we measure the largest excursion
$\Delta_{r_2}=\max_{I_2}{|\delta_{r_2} T|}$ occurring on a
sub-interval $I_2 \subset I_1$ of length $r_2$.  If the mature fronts
dominate the statistics, as in the first scenario, a one-dimensional
section of the scalar field then would look like a sequence of steps.
The distribution of $\Delta_{r_2}$ should then be sharply peaked
around $\delta_{r_1} T $ for any $r_2$. On the contrary, if mature
fronts have a relatively little probability with respect to non-mature
ones, the typical value for $\Delta_{r_2}$ moves to smaller and
smaller values while decreasing $r_2$.  In Figure~\ref{f6} we show
that the second possibility is realized, and thus non-mature fronts
dominate the statistics of the large excursions.
\narrowtext 

Summarizing, saturation comes from a self-similar process and the
signature of the exponent $\zeta_{\infty}$ is felt throughout the
inertial range down to the dissipative scales. In this sense, both
mature and non-mature structures carry the imprinting of saturation:
indeed the dimensions of the set of mature and of the set of
non-mature fronts are equal to $2-\zeta_{\infty}$.  Nevertheless,
since the appearance of a mature front is a relatively rare event, the
dominant contribution to structure functions is due to non-mature
cliffs.

\section{Plateaux}
\label{plat}

\begin{figure} 
\epsfxsize=8truecm 
\epsfbox{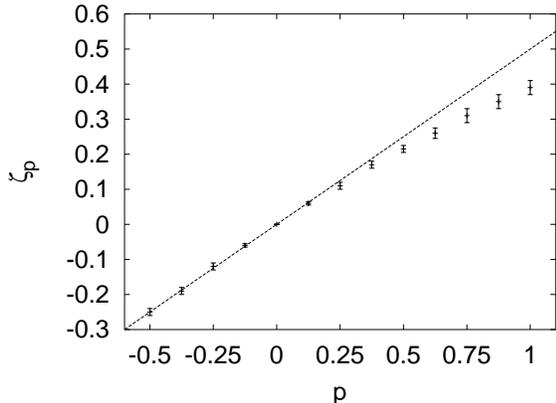}
\caption{Scaling exponents of structure functions for low, 
fractional orders, in the random-forced case. 
Moments are taken with absolute values. The dotted line
is $p/2$.} 
\label{f7} \end{figure}

\begin{figure} 
\epsfxsize=8truecm 
\epsfbox{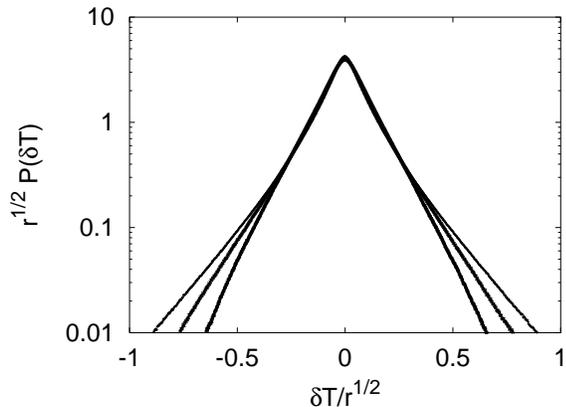}
\caption{Probability density functions of scalar increments for
three separations within the inertial range, rescaled by $r^{1/2}$.}
\label{f8} \end{figure}

Fronts separate regions of space where a very efficient mixing takes
place, and fluctuations of the scalar field are small.  To investigate
the statistical signature of these {\em plateaux} we computed the
low-order moments of scalar increments, shown in Fig.~\ref{f7}. We
observe that the scaling exponents of negative order display an almost
linear dependence on the order, close to $p/2$.  Thus, the scalar
field within the plateaux {\em is not smooth} but it is characterized by
a H\"older exponent 1/2.  This would entail a single rescaling factor,
$r^{1/2}$, for the inner core of the pdf of scalar increments, which
seems to be the case, as in Fig.~\ref{f8}.  This result suggests that
the nature of scalar turbulence within the plateaux could be
understood in simple terms, although, to our knowledge, no definite
explanation of these observations is available up to now.
\narrowtext 
\narrowtext 

\section{Scalar Dissipation}
\label{dissip}
\begin{figure} 
\epsfxsize=7.6truecm 
\epsfbox{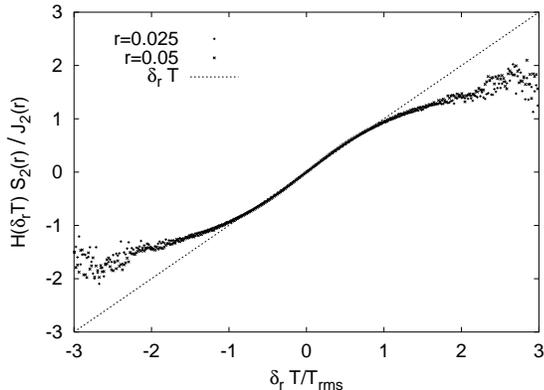}
\caption{The conditional dissipation $H$ 
{\it vs} $\delta_rT$ for two different
separations $r$ in the inertial range.}
\label{f2} \end{figure}

\begin{figure} 
\epsfxsize=7.6truecm 
\epsfbox{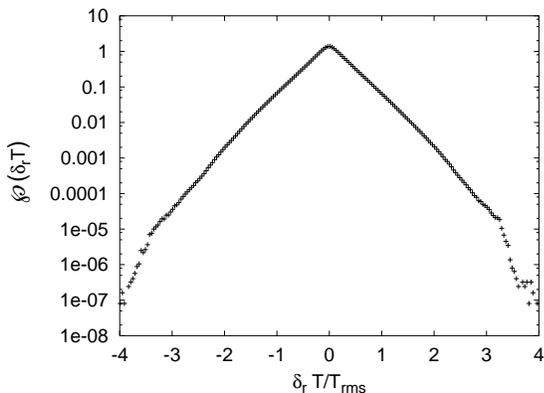}
\caption{The scalar increment pdf for a separation within the 
inertial range.}
\label{f3} \end{figure}

Let us focus our attention on the passive scalar equation (\ref{fp}),
in the presence of a random forcing (\ref{fcorr}).  Denoting $\delta_r
T\equiv T(\bbox{x})-T(\bbox{x}')$ the scalar increment over the
separation $\bbox{r}=\bbox{x}-\bbox{x}^{\prime}$, it is easy to obtain
\cite{K94} the equation for the even-integer structure functions
$S_{2n}$ (odd-integer structure functions are trivially vanishing for
isotropic injection mechanisms):
\begin{equation}
\frac{\partial S_{2n}(r)}{\partial t} + \bbox{\nabla}\cdot\langle\bbox{W}
(\delta_r T)^{2n}\rangle = J_{2n}(r)
\label{general}
\end{equation}
where 
\begin{equation}
\label{gei2n}
J_{2n}\equiv 2n \langle (\delta_r T)^{2n-1} H\rangle ,
\end{equation}
and
\begin{equation}
\label{condW}
\bbox{W}\equiv\langle\left[\bbox{v}(\bbox{x},t)-
\bbox{v}(\bbox{x}',t)\right]|\delta_r T\rangle ,
\end{equation}
\begin{equation}
\label{condH}
H\equiv\langle
\kappa\left(\Delta+\Delta'\right)\delta_r T|\delta_r T\rangle
\end{equation}
are the ensemble average of velocity increment and dissipation
conditioned on the scalar increment $\delta_r T$, respectively.  The
Laplace operators in (\ref{condH}) are evaluated at $\bbox{x}$ and
$\bbox{x}^{\prime}$ respectively.  In our DNS simulations we replaced
the Laplace operator in Eq.~(\ref{fp}) with a bi-Laplacian dissipative
term. The definition of the conditional dissipation $H$ is thus\,: $
H(\delta_rT)\equiv - \kappa\langle
\left(\Delta^2+(\Delta^{\prime})^2\right) \delta_r
T|\delta_rT\rangle$.  The choice of a laplacian or a bilaplacian is
immaterial to all the forthcoming arguments.  In isotropic conditions
the vector $\bbox{W}$ has the form $\bbox{W}= W\hat{\bbox{r}}$, and is
even under a sign reversal of the scalar field, whereas $H$ is odd.
The information carried by the set of equations (\ref{general}), which
hold for any $n$, can be expressed in a single equivalent equation for
the pdf of scalar differences
\begin{equation}
\partial_t {\cal P}(\delta_r T)+ 
\bbox{\nabla} \cdot \left( \bbox{W} {\cal P}(\delta_r T) \right)= 
- {\partial \over \partial \delta_rT} \left( H {\cal P}(\delta_r T) \right) \;.
\label{pdfdT}
\end{equation}
In the statistically stationary state this equation provides a dynamic
link among the three quantities $W,H,{\cal P}$. 
Let us give a few examples which will be relevant to our subject:
$(i)$, if ${\cal P}$ is exponential 
and $H$ is linear, then $W$ is linear; 
$(ii)$, if ${\cal P}$ is exponential and $H$ is constant, then
then $W$ is constant; 
$(iii)$, if ${\cal P}$ is Gaussian
and $W$ is constant, then $H$ is inversely proportional to the 
scalar fluctuation.
\narrowtext \begin{figure} 
\epsfxsize=7.6truecm 
\epsfbox{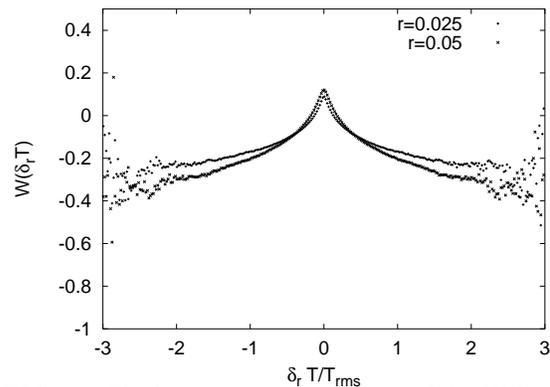}
\caption{The longitudinal component of the \bbox{W} function {\it vs}
$\delta_rT$ for two different values of $r$ in the inertial
range.}
\label{f1} \end{figure}
Let us now look in detail at the numerical results:
$W$ is shown in Fig.~\ref{f1}. 
The negative part for
large values of $\delta_rT$ is physically intuitive\,: the formation
of strong cliffs amounts to the build-up of intense 
scalar gradients, and takes place whenever the 
gradient alignes with the eigenvector of the strain with a negative eigenvalue.
In other words, to obtain a large fluctuation it is necessary that the flow
favors the approaches of elements
of fluid that were initially far apart and characterized
by very different values of the scalar field. 
Conversely, the positive peak for small scalar fluctuations shows that
the ``plateaux'' are regions of extremely efficient mixing, dominated by
small-scale disordered velocity fluctuations with high
turbulent diffusivity.\\
The conditional dissipation function $H({\delta_rT})$ is
shown in Fig.~\ref{f2}. The curve closely follows the linear ansatz 
\cite{K94} up
to values comparable to $T_{rms}$.  For larger values the curve is
however distorted and has a definite tendency to flatten out.
\narrowtext 
As discussed in Sec.~\ref{int-satu}, the tails of scalar increment pdf's
${\cal P}(\delta_rT)$ for various separations $r$ in the inertial
range all collapse onto a single curve\,: ${\cal
P}(\delta_rT)=r^{\zeta_{\infty}}Q({\delta_rT\over
T_{rms}})$.  The curve $r^{-\zeta_{\infty}}{\cal P}(\delta_rT)$
is shown in Fig.~\ref{f3} for one separation in the inertial
range. Its tails (and those of $Q$) decay roughly as an
exponential. 
\narrowtext 

The one-point pdf was shown in Fig.~\ref{f4} and clearly decays as a
Gaussian in the random-forced case.
As a consequence, the exponential tails observed in
Fig.~\ref{f3} cannot continue for ever 
and a sharper than Gaussian fall 
off (see the arguments of Sec.~\ref{dissip})
is expected to set in.  

 From the comparison of the graphs of $W$,$H$ and ${\cal P}$ it
appears that for fluctuations small compared to $T_{rms}$ we are under
the conditions $(i)$, with an almost linear $H$, an approximately
exponential behavior for ${\cal P}$ and a $W \sim |\delta_r T|$.  This
regime is characteristic of ``plateaux'' where efficient mixing occurs
($W>0$) and dissipation is proportional to the intensity of the
fluctuation.  For larger values $\delta_r T$ of the order of a few
$T_{rms}$ we can recognize the case $(ii)$, characterized by the
presence of fronts. Here, the dissipation is not anymore proportional
to the scalar excursion, rather tending to be independent of it. It
has to be remarked that, although large gradients may develop within a
front, they have a negligible statistical weight, so that the main
contribution to the dissipation comes from relatively small
fluctuations of the order
$\kappa^{1/4}\epsilon_{\theta}^{1/2}\epsilon_v^{-1/4}$.  The regime
$(iii)$, where the strength of fluctuations is dominated by the
one-point pdf, is out of reach for our statistics: indeed, a rough
estimate of the amplitudes where this regime sets in is $10\,T_{rms}$.

When the velocity field is rapidly changing in time, the expression for
$W$ can be derived analytically, a fact which allows for 
further considerations
which are presented in Appendix~\ref{diss_kr}.
\section{Scalar fluxes}
\label{fluxes}
\begin{figure} 
\epsfxsize=7.6truecm 
\epsfbox{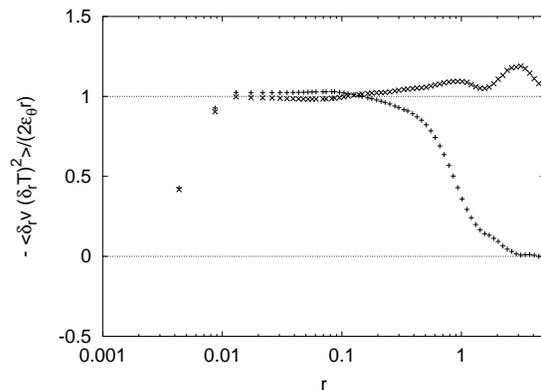}
\caption{The Yaglom's law for the shear-driven ($\times$) and the 
random-forced ($+$) case.}
\label{f-yag} \end{figure}

\begin{figure} 
\epsfxsize=7.6truecm 
\epsfbox{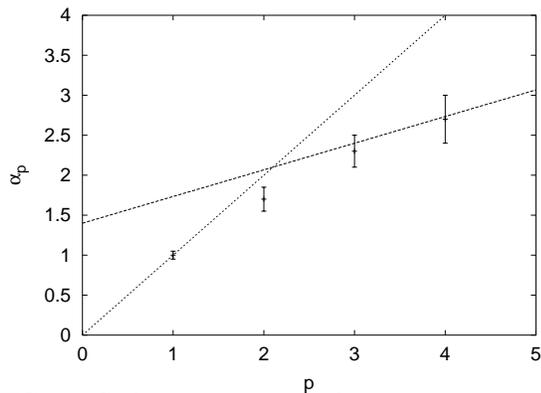}
\caption{Scaling exponents $\alpha_p$ for various 
moments of the scalar flux. The dotted line is the dimensional prediction
$\alpha_p=p$ and the solid line is the asymptote 
$\alpha_p=p/3+\zeta_{\infty}$.}
\label{f-flux} \end{figure}

The classical theory of scalar turbulence presented in
Section~\ref{basic} is based on the picture of a  
 cascade of scalar fluctuations from large to small
scales which takes place at a scale-independent rate $\epsilon_{\theta}$.
More quantitatively, the constancy of the scalar flux is expressed
by means of an exact relation, the
Yaglom's law,  which in two dimensions reads
\begin{equation}
\langle \left[ \delta_r \bbox{v} \cdot \hat{\bbox{r}} \right] 
(\delta_r T)^2 \rangle = -2 \epsilon_\theta r \;.
\label{yag}
\end{equation}
for any $r$ within the inertial range of scales.
In the anisotropic case this relation still holds for the projections 
 onto the isotropic sector.
In Fig.~\ref{f-yag} we show the 
velocity-scalar-scalar correlation which appears in (\ref{yag}) 
computed numerically for both the 
mean gradient and the random forced case.\\
\narrowtext 
The statistics of the scalar flux 
displays as well a remarkable intermittency. To quantify it, let us
introduce the fluctuating scalar flux through the scale $r$ as 
$\Pi_r=\delta_r v (\delta_r T)^2$. We consider moments
$\langle |\Pi_r|^p \rangle \sim r^{\alpha_p}$, which have been 
studied in the context of experimental turbulence in Ref.~\cite{cili}. 
As shown in Figure~\ref{f-flux}, the computed
scaling exponents $\alpha_p$ deviate significantly from the dimensional
prediction $\alpha_p=p$.
The behavior of the exponents for large orders can be explained
with the aid of the results of Section~\ref{dissip}. We observed that
the average velocity fluctuation becomes nearly constant 
for large enough scalar fluctuations
(see Fig.~\ref{f1}). In other words, at small scales
we observe a weak correlation between the scalar and
the velocity fluctuations. A very rough estimate, 
based on the assumption that velocity-scalar correlations are
small, gives us $\alpha_p \sim p/3+\zeta_{2p}$ for large
$p$, which is remarkably close to the observed values. 
This approximation suggests
that there will be no saturation for $\alpha_p$, and that asymptotically
$\alpha_p \simeq p/3+\zeta_{\infty}$.
The experimental data reported in Ref.\cite{cili} support this picture.
\narrowtext 
\section{Conclusions and perspectives}
There are several turbulent systems other than passive scalars which
are characterized by the presence of structures\,: ``plumes'' in
turbulent convection, ``vortex sheets'' in shear-driven hydrodynamic
turbulence and ``current sheets'' in magnetohydrodynamics.  It is thus
natural to ask whether some or all of the results obtained for passive
scalars apply to these other systems.\\ {\em Small-scale
anisotropy.\/} A phenomenology similar to the one for the passive
scalar has been identified for Navier--Stokes turbulence in the
presence of a homogeneous shear flow \cite{PS95} and for kinematic
magnetohydrodynamics both in the presence of incompressible
\cite{ALM00} and compressible \cite{AHMM00} velocity
fields. Concerning Navier--Stokes turbulence, an anisotropic component
of the vorticity third-order correlation is found to be independent of
the Reynolds number in the presence of a large-scale shear
\cite{PS95}. A mecahnism in terms of coherent structures has also been
proposed for the emergence of small scale anisotropy \cite{PS95}.
Even if the effect is less pronounced than in the passive scalar case,
it is likely that large-scale anisotropies do not decay down to the
small scales.  The same happens in magnetohydrodynamics turbulence
where it has been found that for incompressible velocity fields the
hyperskewness of the magnetic field diverges at small scales
\cite{ALM00}. No systematic study has been performed in the context of
turbulent convection.\\ {\em Universality.\/} There is good agreement
about the universality, with respect to energy pumping details, of
scaling exponents in homogeneous and isotropic hydrodynamic
turbulence. For anisotropic turbulence it appears that the exponents
in the isotropic sector coincide with those found in isotropic
turbulence \cite{Arad99}. A similar conclusion has been found
analytically for simple models of kinematic magnetohydrodynamics
\cite{ALM00,AHMM00}.  No results are up to now available for the full
magnetohydrodynamics. As for turbulent convection, there the system is
self sustained and does not need external forcing, thus the problem
has to be rephrased in terms of universality with respect to boundary
conditions. We are not aware of any extensive analysis of this issue.
\\ {\em Saturation.\/} The intrinsic instability of vortex sheets
seems to rule out the possibility of a saturation of intermittency for
Navier-Stokes turbulence. Conversely, the extreme stability and
strength of plumes in convective turbulence suggests that saturation
may occur there. Indeed, buoyancy forces may collaborate with
compressional contributions of large scale velocity leading to sharper
and more long-lived fronts than those observed in passive scalar
turbulence. The situation with the magnetohydrodynamics turbulence remains
to be clarified.

 \vspace{0.5cm}

{\bf Acknowledgements.} Helpful discussions with M.~Chertkov,
A.~Fairhall, U.~Frisch, B.~Galanti, R.H.~Kraichnan, V.~Lebedev,
A.~Noullez, J.-F.~Pinton, I.~Procaccia, A.~Pumir and V.~Yakhot are
gratefully acknowledged. We benefited from the hospitality of the 1999
ESF-TAO Study Center. The INFM PRA Turbo (AC), the INFM PA GEPAIGG01
(AM) and the ERB-FMBI-CT96-0974 (AL) contracts are acknowledged.  This
work was supported in part by the European Commission under contract
HPRN-CT-2000-00162 (``Nonideal Turbulence'').  A.C. was supported by a
``H.~Poincar\'e'' CNRS postdoctoral fellowship.  Simulations were
performed at IDRIS (no~991226) and at CINECA (INFM Parallel Computing
Initiative).

\appendix
\section{Dominant and subdominant terms in the structure functions}
\label{so2}

In order to have a better insight on the angular dependence of scalar
structure functions, it is useful to consider a simple example where
the problem can be tackled analytically. This is the case for the
second order structure function in the Kraichnan advection model
\cite{K68,K94}.  In this model the velocity field
$\bbox{v}=\{v_\alpha, \alpha=1,\ldots,d\}$ advecting the scalar is
incompressible, isotropic, Gaussian, white-noise in time; it has
homogeneous increments with power-law spatial correlations and a
scaling exponent $\xi$ in the range $0< \xi< 2$\,:
\begin{eqnarray}
&&\left\langle [v_\alpha (\bbox{r},t)-v_\alpha
(\bbox{0},0)][v_{\beta} (\bbox{r},t)-v_\beta (\bbox{0},0)]
\right\rangle = \nonumber\\ &&\qquad 2 \delta (t) D_{\alpha\beta} (\bbox{r}),
\label{correlations}
\end{eqnarray}
where,
\begin{eqnarray}
D_{\alpha\beta} (\bbox{r})= D_0 r^{\xi}\left[\left(\xi+d-1\right)
\delta_{\alpha\beta}-\xi\,{r_{\alpha}
r_{\beta}\over r^2}\right].
\label{d-tensor}
\end{eqnarray}
The specific form inside the squared brackets follows from the
incompressibility condition on $\bbox{v}$.

As remarked in Sec.~\ref{sec:sat-rhk}, because of the white-in-time
velocity field, this model leads to closed equations for single-time
multiple-space correlation functions such as
$C_{n}\equiv\left\langle\theta(\bbox{r}_1,t)\ldots\theta
(\bbox{r}_n,t)\right\rangle$ at any order $n$ (see, e.g.,
Refs.~\cite{K68,SS94}).

Their general form in the stationary state (i.e.~$\partial_t C_n=0$)
maintained by the large scale gradient $\bbox{g}$ is \cite{SS94}:
\begin{eqnarray}
&&\sum^n_{i\neq j}\left [D_{\alpha\beta}(\bbox{r}_i - \bbox{r}_j) +
\kappa \delta_{\alpha\beta}\right ] \partial^{\alpha}_{r_i}
\partial^{\beta}_{r_j} \langle\theta(\bbox{r}_1)\ldots
\theta(\bbox{r}_n)\rangle \nonumber \\ &=&\sum^n_{i\neq j} g_{\alpha}
g_{\beta} G_{\alpha\beta}(\bbox{r}_i - \bbox{r}_j)
\langle\theta(\bbox{r}_1) \smash{\mathop{\dots\dots}_{\hat{i}\ \
\hat{j}}} \theta(\bbox{r}_n)\rangle \nonumber \\ &-&2 \sum^n_{i\neq
j}g_{\alpha} D_{\alpha\beta}(\bbox{r}_i - \bbox{r}_j)\partial^b_{r_j}
\langle\theta(\bbox{r}_1) \smash{\mathop{\dots\dots}_{\hat{i}}}
\theta(\bbox{r}_n)\rangle,
\label{kraich_aniso}
\end{eqnarray}
with $G_{\alpha\beta}(\bbox{r})=\langle
v_{\alpha}(\bbox{r},t)v_{\beta}(\bbox{0},t)\rangle$ and where the
translational invariance of scalar correlation functions has been
exploited.  For the sake of simplicity, let us focus the attention on
the equation for the two-point correlation function $C_2(r,\phi)$ in
the two-dimensional case, with $\phi$ denoting the angle between the
mean gradient $\bbox{g}$ and the separation $\bbox{r}$. Restricting
ourselves to the inertial range of scales, such equation has the
simple form\,
\begin{eqnarray}
\label{aniso_ope}
&-&\left[\frac{1}{r}\partial_r(r^{\xi+1}\partial_r) \,+\, 
\frac{\xi+1}{r^{2-\xi}}\partial^2_{\phi}\right]\,C_2(r,\phi) \nonumber \\
&=& g^2 - g^2 r^{\xi}\left[(\xi+1) -\xi \cos^2\phi\right].
\end{eqnarray}
Among the possible functional basis through which $C_2(r,\phi)$ can be
decomposed, the SO(2) representation functions \cite{WKT85}
$U^{(l)}(\phi)\equiv \exp(-i\,l\phi)$ turns out to be particularly
useful \cite{ALPP99}.  Indeed, exploiting this representation we have
\begin{equation}
C_2(r,\phi)= \sum_{l=0} C^{(2l)}_2(r) \cos [(2l)\phi],
\label{so2rapp}
\end{equation}
(only even values of $l$ are involved due to the invariance of
$C_2(r,\phi)$ under the transformation $\phi\mapsto \phi+\pi$) and the
equations for each of the projections $C^{(2l)}_2(r)$ are closed and
independent, and can thus be separately solved.  It is easy to show
that the solution for the isotropic component $C^{(0)}_2(r)$ is
simply\,:
\begin{equation}
C^{(0)}_2(r)=
const - \frac{g^2 L^{\xi}}{2(2-\xi)} r^{2-\xi} + \frac{g^2}{4}r^2,
\end{equation}
where the dimensional constant appearing in the right hand side is the
only non-trivial homogeneous solution (zero mode), of the equation for
the second order correlation function. Here, $L$ is the integral scale
of the problem. It then follows that the structure function in the
isotropic sector $l=0$ is proportional to $\frac{g^2
L^{\xi}}{2(2-\xi)} r^{2-\xi} - \frac{g^2}{4}r^2$, i.e. a sum of power
laws. Conversely, the structure function $\langle\left(T(\bbox{r})
-T(\bbox{0})\right)^2\rangle$ for the total field
$T(\bbox{r})=\theta(\bbox{r}) + \bbox{g}\cdot \bbox{r}$ in the
isotropic sector goes as\,:
\begin{equation}
S_2^{(0)}(r)=\frac{g^2L^{\xi}}{2(2-\xi)} r^{2-\xi}.
\end{equation}
This means that the two structure functions, for the fluctuating field
$\theta(\bbox{r})$ and for the total field $T(\bbox{r})$ behave, in
the isotropic sector, in the same way $\propto r^{2-\xi}$, with
decreasing $r$; but while those for $\theta(\bbox{r})$ are sums of
power laws, structure functions of $T(\bbox{r)}$ have a pure power law
scaling. It is then useful to work with the total field $T(\bbox{r})$
to obtain cleaner scaling behaviors.

\section{Scalar dissipation and saturation in the Kraichnan model}
\label{diss_kr}
Let us now specialize the analysis (see also Ref.~\cite{K97})
for the Kraichnan advection model
(\ref{correlations}).
Exploiting homogeneity, incompressibility and 
the $\delta$-correlation in time, $\bbox{\nabla}\cdot\langle\bbox{W}
(\delta_r T)^{2n}\rangle$ can be evaluated exactly and its expression reads
\cite{K94}:
\begin{equation}
\bbox{\nabla}\cdot\langle\bbox{W} (\delta_r T)^{2n}\rangle=
-\frac{2}{r^{d-1}}\frac{\partial}{\partial r}\left(r^{d-1}\eta(r) 
\frac{\partial S_{2n}(r)}{\partial r}
 \right)
\label{avvet}
\end{equation}
where $\eta(r)=D_0 (d-1)/2\,r^{\xi}$ is the relative diffusion operator.
On the contrary, $J_{2n}(r)$ does not involves solely
two-point averages and this means that, in general,  it
cannot be simply expressed in term of structure functions.\\
In the stationary state (i.e.~$\partial S_{2n}/\partial t=0$) 
and for separations, $r$, in the inertial range of scales 
(i.e.~$\eta_{\theta}\ll r\ll L$) 
a power-law behavior for $S_{2n}(r)$ (i.e.~$S_{2n}(r)=s_{2n}r^{\zeta_{2n}}$) 
is expected. 
The same law, with exponent $\zeta_{2n}-\zeta_2$, holds for $J_{2n}$
to ensure the balance between the advective and the diffusive term, 
$J_{2n}$, in Eq.~(\ref{general}). 
Imposing the
balance in Eq.~(\ref{general}) for power law coefficients and using
 (\ref{avvet}), the
following equation is obtained:
\begin{equation}
\zeta_{2n}(d+\zeta_{2n}-\zeta_2)=
d\,\zeta_2 \frac{j_{2n}}{s_{2n}}\frac{s_2}{j_2},
\label{eqbal}
\end{equation}
where we have posed $J_{2n}=j_{2n} r^{\zeta_{2n}-\zeta_2}$.

Let us discuss the consequences of saturation on 
Eq.~(\ref{general}) and thus on (\ref{eqbal}). When scaling exponents
 saturate, for $n\geq\overline{n}$ we have $\zeta_{2n}=\zeta_{\infty}$  
and the left hand side of Eq.~(\ref{eqbal}) does not depend on $n$ any longer. 
To ensure the balance for all $n\geq \overline{n}$, 
the same independence on $n$ thus holds also in the right hand side of 
(\ref{eqbal}), that implies:
\begin{equation}
\frac{j_{2n}}{s_{2n}}\frac{s_2}{j_2}=C\qquad\mbox{or}\qquad 
2n \langle (\delta_r T)^{2n-1} H\rangle =C J_2 \frac{S_{2n}}{S_{2}} 
\label{rel2}
\end{equation}
where $C$ may depends on $\xi$ and/or $d$, but not on $n$.  Notice
that complete saturation, i.e.~$\zeta_{\infty}=\zeta_2$, yields
$C=1$.\\ We now focus on the consequences of (\ref{rel2}) on the
behavior {\it vs} $\delta_r T$ of $H(\delta_r T)$ and $K(\delta_r T)$,
with
\begin{equation}
\label{condK}
K\equiv \kappa
\langle \left|\nabla\delta_r T+\nabla'\delta_r T
\right|^2|\delta_r T\rangle ,
\end{equation}
a conditional mean related to dissipation.\\ To do that, we have to
assume a specific form for the pdf of scalar increments ${\cal
P}(\delta_r T)$. Our attention being focused on values of $\delta_r
T\gtrsim T_{rms}$, so that ${\cal P}(\delta_r T)\propto
r^{\zeta_{\infty}}{\cal Q}\left (\frac{\delta_r T}{T_{rms}}\right )$,
statistical relations involving high-order structure functions
(e.g.~their ratio) are thus fully controlled by ${\cal Q}$.  As a
first choice, fully justified solely for the extreme tails $\delta_r T
\gg T_{rms}$, we can thus assume a Gaussian shape for ${\cal Q}$. When
doing this, recalling that for a Gaussian pdf we have $\langle
(\delta_r T)^{2n} \rangle=(2n-1) \langle (\delta_r T)^{2n-2}\rangle$,
we immediately realize that the second of (\ref{rel2}) is satisfied by
\begin{equation}
\label{h1}
H=C\frac{J_2}{\delta_r T\, S_2} \qquad\mbox{for}\qquad \delta_r T\gg T_{rms}.
\end{equation}
It is now possible to show that, as a consequence of homogeneity alone
(see Ref.~\cite{PC93}) $H$ and $K$ are related by the following
identity:
\begin{equation}
H\, P \equiv \frac{\partial}{\partial \delta_rT}[K\,P],
\label{ident}
\end{equation}
from which, using the Gaussianity of ${\cal Q}$, one obtains that at
large values of $\delta_r T$\,: $K\propto \frac{H}{\delta_r T}$.\\ It
immediately follows according to (\ref{h1}) that the conditional mean
$K$ behaves as
\begin{equation}
K\propto \frac{H}{\delta_r T}\propto \frac{1}{(\delta_r T)^2}.
\label{k1}
\end{equation}
We have already stressed that the behaviors {\it vs} $\delta_r T$ of
the conditional means $H(\delta_r T)$ and $K(\delta_r T)$ depend on
the specific assumption made on the asymptotic shape of ${\cal
Q}$. However, in most experimental situations, one does not have
access to the strongest events controlling the extreme tails of ${\cal
Q}$.  Although one can clearly observe the rescaling (\ref{pdf}) and
thus gather information on the saturation, ${\cal Q}$ might still not
have relaxed on the single point pdf.\\ To better illustrate this
point, let us assume that we can control scalar increments statistics
for $\delta_r T\gtrsim T_{rms}$, but far from the asymptotic situation
$\delta_r T\gg T_{rms}$.  We thus rule out Gaussianity for ${\cal Q}$
and assume a slower decaying, e.g.~of exponential type, as suggested
by the numerical results.  When doing this, recalling that for an
exponential pdf we have $\langle (\delta_r T)^{2n} \rangle=2n \langle
(|\delta_r T|)^{2n-1}\rangle$, we immediately realize that the second
equation in (\ref{rel2}) is now satisfied by
\begin{equation}
\label{h1expo}
H=sign(\delta_r T)\frac{C\,J_2}{S_2}\qquad\mbox{for}\qquad \delta_r T
\gtrsim T_{rms}.
\end{equation}
From the identity (\ref{ident}), it also follows that $K$ is now
$\propto H$.  We can conclude that, at least in the range $\delta_r
T\gtrsim T_{rms}$ (but still far from $\delta_r T\gg T_{rms}$) and for
an exponential decay of ${\cal Q}$ in (\ref{pdf}), saturation is thus
compatible with the behaviors of $H$ and $K$ displaying no dependence
on $\delta_r T$.

\end{document}